\documentclass[12pt]{amsart}

\usepackage[T1]{fontenc}
\usepackage[utf8]{inputenc}

\usepackage{amssymb,latexsym}
\usepackage{enumerate}
\usepackage[mathscr]{eucal}
\usepackage{upref}
\usepackage{esint}
\usepackage{epsfig}
\usepackage{verbatim}
\usepackage{multicol}
\usepackage{multirow}
\usepackage{graphicx, color, psfrag, epstopdf,subfigure}
\usepackage{pstricks,pst-node}
\usepackage{tikz,pgfplots,paralist,mathtools,todonotes,placeins}

\usepackage[paper=a4paper,left=30mm,right=20mm,top=25mm,bottom=30mm]{geometry}

\makeatletter
\@namedef{subjclassname@2010}{
 \textup{2010} Mathematics Subject Classification}
\makeatother

\newtheorem{thm}{Theorem}[section]

\newtheorem{lem}[thm]{Lemma}

\theoremstyle{definition}
\newtheorem{defin}[thm]{Definition}

\numberwithin{equation}{section}

 \newcommand{\N}{\mathbb{N}}
 \newcommand{\R}{\mathbb{R}}

\newcommand{\dokendDef}{\ensuremath{\hfill\square}}

\newcommand{\tcb}{\textcolor{blue}}
\newcommand{\tco}{\textcolor{red}}

\def\vect#1{\mbox{\boldmath$ #1$}}                   

\newcommand{\bA}{{\vect A}}
\newcommand{\ba}{{\vect a}}
\newcommand{\bB}{{\vect B}}

\newcommand{\bC}{{\vect C}}
\newcommand{\bc}{{\vect c}}
\newcommand{\bD}{{\vect D}}
\newcommand{\bd}{{\vect d}}
\newcommand{\be}{{\vect e}}

\newcommand{\bff}{{\vect f}}

\newcommand{\bI}{{\vect I}}
\newcommand{\bl}{{\vect l}}
\newcommand{\bL}{{\vect L}}
\newcommand{\bn}{{\vect n}}

\newcommand{\br}{{\vect r}}
\newcommand{\bR}{{\vect R}}

\newcommand{\bS}{{\vect S}}
\newcommand{\bT}{{\vect T}}
\newcommand{\bt}{{\vect t}}
\newcommand{\bu}{{\vect u}}
\newcommand{\bU}{{\vect U}}

\newcommand{\bw}{{\vect w}}
\newcommand{\bx}{{\vect x}}
\newcommand{\bX}{{\vect X}}

\newcommand{\bY}{{\vect Y}}
\newcommand{\bz}{{\vect z}}
\newcommand{\bZ}{{\vect Z}}

\newcommand{\bone}{{\vect 1}}
\newcommand{\bzero}{{\vect 0}}

\newcommand{\bLambda}{{\vect \Lambda}}

\newcommand{\ou}{{\overline u}}

\newcommand{\ow}{{\overline w}}

\newcommand{\olambda}{{\overline \lambda}}

\newcommand{\obA}{{\overline \bA}}
\newcommand{\oba}{{\overline \ba}}
\newcommand{\obC}{{\overline \bC}}
\newcommand{\obc}{{\overline \bc}}
\newcommand{\obD}{{\overline \bD}}
\newcommand{\obd}{{\overline \bd}}
\newcommand{\obe}{{\overline \be}}

\newcommand{\obI}{{\overline \bI}}

\newcommand{\obL}{{\overline \bL}}
\newcommand{\obn}{{\overline \bn}}

\newcommand{\obt}{{\overline \bt}}
\newcommand{\obT}{{\overline \bT}}
\newcommand{\obu}{{\overline \bu}}

\newcommand{\obw}{{\overline \bw}}
\newcommand{\obx}{{\overline \bx}}
\newcommand{\obz}{{\overline \bz}}

\newcommand{\op}{{\overline p}}
\newcommand{\oq}{{\overline q}}

\def\vects#1{\mbox{\scriptsize \boldmath$ #1$}}

\newcommand{\ibn}{{\vects n}}
\newcommand{\ibt}{{\vects t}}

\newcommand{\ibw}{{\vects w}}

\newcommand{\iobn}{{\overline{\vects n}}}
\newcommand{\iobt}{{\overline{\vects t}}}

\newcommand{\iobw}{{\overline{\vects w}}}

\newcommand{\diag}{\mbox{diag}}

\definecolor{pumpkin}{rgb}{1.0, 0.46, 0.09}
\definecolor{cadmiumgreen}{rgb}{0.0, 0.42, 0.24}

\definecolor{darkmagenta}{rgb}{0.55, 0.0, 0.55}

\begin{document}
\baselineskip=17pt

\title[Radially symmetric solutions]
{Radially symmetric solutions\\ of the ultra-relativistic Euler equations\\ in several space dimensions}

\author[M. Kunik]{Matthias Kunik}
\address{Otto-von-Guericke-Universit\"{a}t Magdeburg\\
Institut f\"ur Analysis und Numerik\\
Universit\"{a}tsplatz 2 \\
D-39106 Magdeburg \\
Germany}
\email{matthias.kunik@ovgu.de}

\author[A. Kolb]{Adrian Kolb}
\address{IGPM, RWTH Aachen University\\
Templergraben 55\\
D-52056 Aachen\\
Germany}
\email{kolb@eddy.rwth-aachen.de}

\author[S. M\"uller]{Siegfried M\"uller}
\address{IGPM, RWTH Aachen University\\
Templergraben 55\\
D-52056 Aachen\\
Germany}
\email{mueller@igpm.rwth-aachen.de}

\author[F. Thein]{Ferdinand Thein}
\address{IGPM, RWTH Aachen University\\
Templergraben 55\\
D-52056 Aachen\\
Germany}
\email{thein@igpm.rwth-aachen.de}

\date{\today}

\begin{abstract}
    The ultra-relativistic Euler equations for an ideal gas are described in terms of the pressure,
    the spatial part of the dimensionless four-velocity and the particle density.
    Radially symmetric solutions of these equations are studied in two and three space dimensions.
    Of particular interest in the solutions are the formation of shock waves and a pressure blow up.
    For the investigation of these phenomena we develop a one-dimensional scheme using
    radial symmetry and integral conservation laws.
    We compare the numerical results with solutions of multi-dimensional high-order numerical schemes for general initial data in two space dimensions.
    The presented test cases and results may serve as interesting benchmark tests for multi-dimensional solvers.
\end{abstract}



\maketitle
%
%
\section{Introduction}
\label{intro}
In this paper we focus on radially symmetric solutions of a special rela\-tivistic system which is much simpler than flows in general relativistic theory.
Interestingly, even compared to the classical Euler equations of non-relativistic gas dynamics the equations we consider exhibit a simpler mathematical structure.\\
We are concerned with the ultra-relativistic equations for a perfect fluid in Minkowski space-time $\mathbf{x}=(x_1,x_2,x_3)$, $t=x_0$, namely
\begin{equation}\label{div}
    \sum_{\beta=0}^{3}\frac{\partial T_{\alpha \beta}}{\partial x_{\beta}} = 0
\end{equation}
with $\alpha,\beta\in\{0,1,2,3\}$ and the energy-momentum tensor
\begin{equation*}
    T_{\alpha\beta} =-\tilde{p} g_{\alpha\beta}+4\tilde{p}u_\alpha u_\beta
\end{equation*}
for the ideal ultra-relativistic gas. Here $\tilde{p}$ represents the pressure,
$\mathbf{u} \in \R^3$ is the spatial part of the four-velocity vector $(u_0,u_1,u_2,u_3)=(\sqrt{1+|\mathbf{u}|^2},\mathbf{u})$.
The flat Minkowski metric is given as
\begin{equation*}
    g_{\alpha \beta} = \begin{cases}
        +1, &\alpha = \beta = 0,\\
        -1, &\alpha = \beta = 1,2,3,\\
        \hphantom{-}0, &\alpha \neq \beta.
    \end{cases}
\end{equation*}
We note that the quantities $u_{\alpha}$, $T_{\alpha \beta}$, $g_{\alpha \beta}$ and even $x_{\alpha}$ are usually 
written as Lorentz-invariant tensors with \textit{upper indices}
instead of  lower indices in order to make use of Einstein's summation convention. But in the following calculations these upper indices could be mixed up with powers.
Since we will not make use of the lowering and raising of Lorentz-tensor indices, our change of the notation will not lead to confusions.
For the physical background we refer to Weinberg \cite[Part I, pp 47-52]{Weinberg}, further details can be found in Kunik \cite[Chapter 3.9]{Kunikthesis}
and for the corresponding classical Euler equations see Courant and Friedrichs \cite{CF}.
For a general introduction to the mathematical theory of hyperbolic conservation laws see Bressan \cite{Bressan} and Dafermos \cite{CD}.
An overview of radially symmetric solutions to conservation laws
is given in the survey paper by Jenssen \cite{HJ}.
Previous results on the numerical treatment of the ultra-relativistic Euler equations are given in Abdelrahman et al. \cite{Abdelrahman3} proposing a front tracking scheme and for kinetic schemes we cite Kunik et al. \cite{KQW1,KQW2,KQW3}.
For a recent treatment of the ultra-relativistic equations, especially in the context of symmetric hyperbolic systems,
we refer to Freist\"uhler \cite{HF2019}, Ruggeri and Masaru \cite{RM2021} and the references therein.
The outline of the remaining paper is as follows. In Section \ref{sec:eqns} we present the equations subject of study in this work. In Section \ref{scheme1} the one--dimensional scheme to compute the radially symmetric solutions is given.
In Section \ref{sec:numeric} we first validate the one-dimensional scheme in case of self-similar solutions where we compare with solutions of an ODE system. We further present three additional radially symmetric benchmark problems without scale invariance. With these examples we verify that the one-dimensional solver can be used to validate genuinely multi-dimensional solvers. A conclusion is given in Section \ref{sec:conclusion}. For the readers convenience we also provide the eigenstructure of the system under consideration in Appendix \ref{app:eig_sys_ure} which, to the best of our knowledge, cannot be found in the literature.
\section{Conservative Formulations of the Equations}\label{sec:eqns}
In the following we introduce two kind of conservative formulations of \eqref{div}. First, the multi--dimensional form  of the ultra--relativistic Euler equations and second the conservative radially symmetric form.
In both cases we consider either $d = 2$ or $d = 3$ space dimensions. Then the unknown quantities $\tilde{p}$ and $\mathbf{u} = (u_1,\ldots,u_d) \in \R^d$
satisfying \eqref{div}
depend on time $t\geq 0$ and position $\mathbf{x} = (x_1,\ldots,x_d) \in \R^d$.\\
Putting $\alpha = 0$ in \eqref{div} gives the conservation of energy
\begin{equation}\label{energy_general}
    \frac{\partial}{\partial t}\left(3\tilde{p} + 4\tilde{p}|\mathbf{u}|^2 \right) + \sum\limits_{k=1}^d\frac{\partial}{\partial x_k}\left(4\tilde{p}u_k\sqrt{1 + |\mathbf{u}|^2} \right)=0,
\end{equation}
whereas for $\alpha = j = 1,\dots,d$ we obtain the conservation of momentum
\begin{equation}\label{momentum_general}
    \frac{\partial}{\partial t}\left(4\tilde{p}u_j \sqrt{1 + |\mathbf{u}|^2} \right) + \sum\limits_{k=1}^d\frac{\partial}{\partial x_k}\left(\tilde{p}\delta_{jk} + 4\tilde{p}u_j u_k\right)=0,
    \quad j = 1,\dots,d.
\end{equation}
In this paper, we study radially symmetric solutions and construct corresponding schemes to solve the ultra-relativistic Euler equations \eqref{energy_general},
\eqref{momentum_general} in two and three space dimensions. 
Here we focus on the case $d=2$, since a detailed treatment of the case $d=3$ is presented in \cite[Sec. 2, Eqn. (2.5)]{KLW}. For completeness we give a summary of the radially symmetric equations for $d=2,3$ in \eqref{radsys} in the present paper.\\
Assume for a moment a smooth solution $\tilde{p}, \mathbf{u}$ of the ultra-relativistic Euler equations \eqref{energy_general}, \eqref{momentum_general}.
We put $r = |\mathbf{x}|$ for $r > 0$ and look for radially symmetric solutions
\begin{equation}\label{rad}
    p = p(t,r) = \tilde{p}(t,\mathbf{x}) > 0\,, \quad \mathbf{u}(t, \mathbf{x}) = \frac{u(t,r)}{r}\,\mathbf{x}.
\end{equation}
Here the quantity $\mathbf{u}(t, \mathbf{x}) \in \R^2$ is completely determined by a new 
\textit{real-valued quantity} $u(t,r)$ depending on $t > 0$, $r > 0$.
For continuity we have the boundary condition
\begin{equation}\label{zeroboundary}
    \lim \limits_{r \searrow 0} u(t,r)=0, \quad t > 0.
\end{equation}
Note that $\mathbf{n} = \frac{1}{r}\mathbf{x}$ is the outer normal vector field of the circle $\partial \mathcal{B}_R$ bounding the ball
\[
  \mathcal{B}_R=\{\mathbf{x} \in \R^2\,:\,|\mathbf{x}| \leq R \}
\]
of radius $R>0$ and that $|\mathbf{u}|^2=u^2$ as well as $u=\mathbf{u}\cdot \mathbf{n}$.
Therefore, it is natural to apply the Gaussian divergence theorem for the integration of 
the divergence term in \eqref{energy_general} over $\mathcal{B}_R$ to make
use of the radial symmetry of the solutions. 
We obtain with \eqref{rad} for any fixed $R>0$
\begin{equation*}
    2\pi \frac{\partial}{\partial t}\int \limits_0^R\left(3p(t,r)+4p(t,r)u^2(t,r)\right) r\, dr + \int \limits_{\partial \mathcal{B}_R}4pu\sqrt{1 + u^2}\,ds = 0.
\end{equation*}
The contour integral with the differential length $ds$ on the right-hand side is constant. Hence we have
\begin{equation}\label{energy_int}
    \begin{split}
        &\frac{\partial}{\partial t}\int \limits_0^R\left(3p(t,r) + 4p(t,r)u^2(t,r)\right) r\, dr\\
        &+ 4p(t,R)u(t,R)R\sqrt{1 + u^2(t,R)} = 0.
    \end{split}
\end{equation}
This idea does not work for the momentum equation \eqref{momentum_general}, because \eqref{rad} would give zero after integration 
over
$\mathcal{B}_R$.
Here we integrate \eqref{momentum_general} for $j=2$ over the upper half-ball
\[
  \mathcal{B}^+_R = \{\mathbf{x}=(x_1,x_2) \in \R^2\,:\,x_2 \geq 0\},
\]
use the Gaussian divergence theorem and polar coordinates
\begin{equation*}
    x_1 = r\cos{\varphi},\; x_2 = r\sin{\varphi}
\end{equation*}
with $0<r<R$ and $0 < \varphi <\pi$ and obtain with \eqref{rad}
\begin{equation}\label{momentum_int}
    \begin{split}
        &2 \frac{\partial}{\partial t}\int \limits_0^R 4p(t,r)u(t,r)r\sqrt{1+u^2(t,r)}\, dr\\
        &+2R\left(4p(t,R)u^2(t,R)+p(t,R)\right) - 2\int \limits_0^R p(t,r)dr = 0.
    \end{split}
\end{equation}
Now we differentiate the Eqns.~\eqref{energy_int}, \eqref{momentum_int}
with respect to $R>0$. Afterwards we replace $R$ by the better suited variable $x>0$.\\
We put $p=p(t,x)$, $u=u(t,x)$ for abbreviation and have the $2$ by $2$ system
\begin{equation}\label{ultra_rad1}
    \begin{dcases}
        \frac{\partial}{\partial t}\left(x p (3 + 4u^2) \right) + \frac{\partial}{\partial x}\left(4 x p u\sqrt{1 + u^2} \right) &= 0,\\
        \frac{\partial}{\partial t}\left(4 x p u\sqrt{1 + u^2} \right) + \frac{\partial}{\partial x}\left(x p (1 + 4u^2) \right) &= p.
    \end{dcases}
\end{equation}
The validity of this system may also be checked by differentiation from \eqref{energy_general}, \eqref{momentum_general} and \eqref{rad}.
The solutions of \eqref{ultra_rad1} are restricted to the state space $\mathcal{S}_{eul}=\{(p,u) \in \R^2\,:\,p > 0\}$.\\
It is well-known that even for smooth initial data, where the fields are prescribed at $t=0$, the solution may develop shock discontinuities.
This requires a weak form of the conservation laws in \eqref{ultra_rad1}.\\
For the formulation of weak solutions we first introduce a transformation in state space. With
\[
  \tilde{\mathcal{S}}_{eul} = \{(a,b) \in \R^2\,:\,|b| < a\}
\]
there is a one-to-one transformation $\Theta: \mathcal{S}_{eul} \mapsto \tilde{\mathcal{S}}_{eul}$ given by
\begin{equation}\label{statetrans}
    \Theta(p,u) = \begin{pmatrix} p(3 + 4u^2)\\ 4pu\sqrt{1 + u^2}\\ \end{pmatrix}
                = \begin{pmatrix} a\\ b\\ \end{pmatrix}.
\end{equation}
The inverse transformation is given by
\begin{equation}\label{inverse_statetrans}
    p = \frac{1}{3}\left( \sqrt{4a^2 - 3b^2} - a \right),\quad u = \frac{b}{\sqrt{4p(p + a)}}.
\end{equation}
In \cite{Kunikthesis} we have used contour integrals for weak solutions of conservation laws, following Oleinik's formulation \cite{Oleinik:1957} for a scalar conservation law.
We put
\begin{equation}\label{cdefinition}
    c = c(a,b) = \frac{5}{3}a - \frac{2}{3}\sqrt{4a^2 - 3b^2},
\end{equation}
and obtain especially for a smooth solution $a$, $b$ and for each convex domain $\Omega \subset Q$ with piecewise smooth boundary $\partial \Omega \subset Q$:
\begin{equation}\label{weak_contour}
    \begin{array}{ll}
        \int\limits_{\partial \Omega} x a\,dx - x b\,dt = 0,\quad \int\limits_{\partial \Omega} x b\,dx - x c\,dt = \dfrac12 \iint \limits_{\Omega} (a - c)\,dtdx.
    \end{array}
\end{equation}
This is a proper weak formulation which will be used next for more general piecewise smooth solutions.
Using the transformation in state space \eqref{statetrans} we obtain an initial value problem for $a$ and $b$. In the quarter plane $t>0$, $x>0$ we have to require that $|b(t,x)|<a(t,x)$.
Then we prescribe for $x>0$ the \textit{two initial functions}
\begin{equation}\label{initial_final}
    \lim \limits_{t \searrow 0}a(t,x) = a_0(x),\quad \lim \limits_{t \searrow 0}b(t,x) = b_0(x),\; x > 0
\end{equation}
with $|b_0(x)|<a_0(x)$ for $x>0$.\\
%
In the presence of shock waves we also obtain a very simple characte\-rization of the entropy condition, see \cite[Chapter 2.1]{Abdelrahman1} for more details:
If for $p_-,p_+>0$ the left state $(p_-,u_-)$ can be connected to the right state $(p_+,u_+)$ by a single shock satisfying the Rankine-Hugoniot jump conditions,
then this shock wave satisfies the correct entropy condition if and only if $u_- > u_+$. This condition can also be checked easily for our numerical solutions with shock curves.\\
In Section \ref{scheme1} the conservation laws \eqref{weak_contour} are used in order to develop a one--dimensional numerical scheme to compute the radially symmetric solutions of the system \eqref{energy_general}, \eqref{momentum_general} in two space dimensions. Compared to the corresponding scheme for $d=3$ presented in \cite{KLW} it turns out that both schemes have essentially the same structure. To be more precise, for $d=2$ only the subroutine ``Euler'' defined in Def. \ref{eulerfunction} is slightly modified.

\section{Formulation of a numerical scheme for the radially symmetric solutions}\label{scheme1}
\begin{figure}[h!]
    \center
    \begin{tikzpicture}
        %
        \draw[thick,->] (0,0) -- (0,5.5);
        \draw[thick,->] (0,0) -- (5.5,0);
        \node at (-0.2,5.5) {$x$};
        \node at (5.5,-0.3) {$t$};
        \node at (0.9,2.2) {$\mathcal{D}$};
        \node at (-0.2,-0.2) {$0$};
        \node at (-0.4,2.6) {$x_*$};
        \node at (-0.8,4.6) {$x_{*}+\lambda t_*$};
        \node at (0.9,2.2) {$\mathcal{D}$};
        \node at (2,-0.3) {$t_*$};
        \fill[black] (0,4.6) circle (2pt);
        \fill[black] (2,0) circle (2pt);
        \fill[black] (0,0) circle (2pt);
        \fill[black] (2,2.6) circle (1pt);
        \draw[very thick] (0,0)--(2,0)--(2,2.6)--(0,4.6)--(0,0);
        \draw (0,2.6)--(2,2.6);
    \end{tikzpicture}
    \caption{The computational domain $\mathcal{D}$}
    \label{fig_domain}
\end{figure}
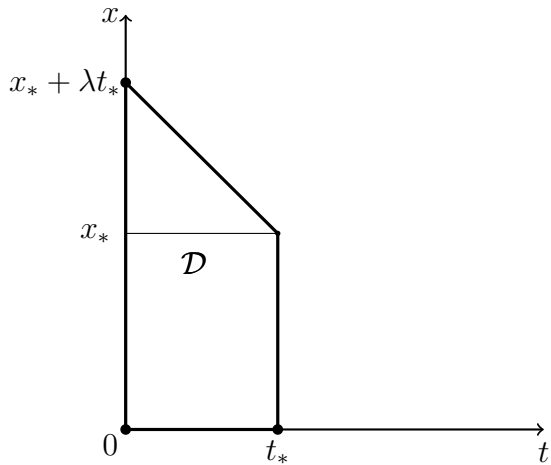

\begin{figure}[h!]
    \begin{tikzpicture}
        \draw[thick] (0,2) -- (0.6,2);
        \fill[black] (0.8,2) circle (2pt);
        \fill[black] (1,2) circle (2pt);
        \fill[black] (1.2,2) circle (2pt);
        \draw[thick] (1.4,2) -- (2,2);
        \draw[thick] (0,4) -- (0.6,4);
        \fill[black] (0.8,4) circle (2pt);
        \fill[black] (1,4) circle (2pt);
        \fill[black] (1.2,4) circle (2pt);
        \draw[thick] (1.4,4)-- (2,4);
        \draw[thick] (0,6) -- (0.6,6);
        \fill[black] (0.8,6) circle (2pt);
        \fill[black] (1,6)circle (2pt);
        \fill[black] (1.2,6) circle (2pt);
        \draw[thick] (1.4,6) -- (2,6);
        \node at (-0.2,-2.3) {$0$};
        \node at (-1,2) {$\overline{x}-\Delta x/2$};
        \fill[black] (4,6) circle (2pt);
        \node at (-1,6) {$\overline{x}+\Delta x/2$};
        \node at (-0.3,4) {$\overline{x}$};
        %
        \draw[thick] (0,-2) -- (0.6,-2);
        \fill[black] (0.8,-2) circle (2pt);
        \fill[black] (1,-2) circle (2pt);
        \fill[black] (1.2,-2) circle (2pt);
        \draw[very thick,-stealth] (1.4,-2) -- (7,-2);
        \node at (7,-2.3) {$t$};
        \draw[thick] (0,-2) -- (0,-1.4);
        \fill[black] (0,-1.2) circle (2pt);
        \fill[black] (0,-1) circle (2pt);
        \fill[black] (0,-0.8) circle (2pt);
        \draw[very thick,-stealth] (0,-0.6) -- (0,8.5);
        \node at (-0.3,8.4) {$x$};
        \draw[thick] (2,-2) -- (2,-1.4);
        \fill[black] (2,-1.2) circle (2pt);
        \fill[black] (2,-1) circle (2pt);
        \fill[black] (2,-0.8) circle (2pt);
        \draw[thick] (2,-0.6) -- (2,4);
        \draw[thick] (4,-2) -- (4,-1.4);
        \fill[black] (4,-1.2) circle (2pt);
        \fill[black] (4,-1) circle (2pt);
        \fill[black] (4,-0.8) circle (2pt);
        \draw[thick] (4,-0.6) -- (4,2);
        \draw[thick] (6,-2) -- (6,-1.4);
        \fill[black] (6,-1.2) circle (2pt);
        \fill[black] (6,-1) circle (2pt);
        \fill[black] (6,-0.8) circle (2pt);
        \draw[thick] (6,-0.6) -- (6,0);
        \draw[thick] (2,4) -- (6,4);
        \draw[thick] (2,2) -- (4,2);
        \draw[thick] (2,6) -- (4,6);
        \draw[thick] (4,2) -- (4,6);
        \node at (2,-2.3) {$\overline{t}-\Delta t$};
        \node at (4,-2.3) {$\overline{t}$};
        \node at (6,-2.3) {$\overline{t}+\Delta t$};
        \draw[thick] (2,4) -- (6,8);
        \draw[thick] (2,4) -- (6,0);
        \draw[thick] (6,0) -- (6,8);
        \draw[very thick,-stealth] (4,2) -- (5,1);
        \node at (4.8,0.8) {$\gamma_2$};
        \draw[thick,-stealth] (4,6) -- (3,5);
        \node at (2.72,5.2) {$\gamma_1$};
        \draw[very thick,-stealth] (6,4) -- (6,6);
        \node at (6.4,6) {$\gamma_0$};
        \node at (4.5,4.3) {$\Omega$};
        \node at (4.8,5.9) {$(a_+,b_+)$};
        \node at (4.8,2.1) {$(a_-,b_-)$};
        \node at (6.6,4) {$(a',b')$};
    \end{tikzpicture}
    \caption{The balance region $\Omega$}
    \label{figeul}
\end{figure}
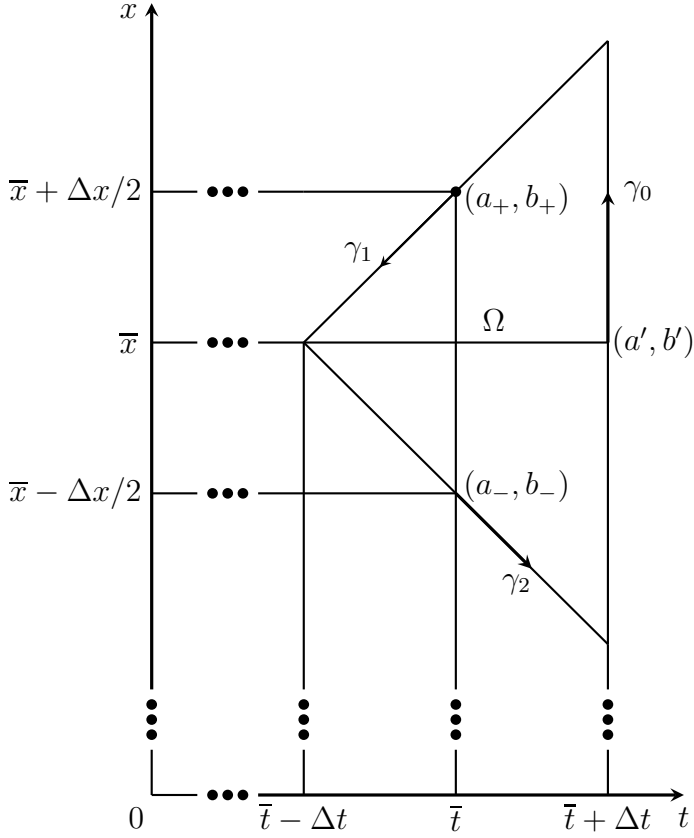

We develop a one-dimensional numerical scheme for the initial value problem of
the radially symmetric ultra-relativistic Euler equations in two space dimensions.
As mentioned earlier the details for the three-dimensional case can be found in \cite{KLW}.
We show that our scheme preserves positive pressure.
The method of contour-integration for the formulation of the balance laws \eqref{weak_contour} is used to construct a function called ``Euler''. 
This function enables us to obtain the time evolution of the numerical solution on a staggered grid.
More precisely it allows us to construct the solution $(a',b')$ at the next time step
from the solution $(a_{\pm},b_{\pm})$ in two neighboring grid points at the previous time step according to Figure \ref{figeul}.
Parts of the construction are exactly the same as in \cite[Section 4]{KLW}, namely the determination of the grid points. It finally turns out that only the routine ``Euler''
has to be modified for the solution of the two-dimensional model. For the sake of a better understanding we now present the detailed construction.
First we determine the computational domain and define some quantities which are needed for its discretization.
\begin{enumerate}[1)]
    \item Given are $t_*,x_*>0$ in order to calculate a numerical solution of the initial value problem \eqref{weak_contour}, \eqref{initial_final}
          in the time range $[0,t_*]$ and the spatial range $[0,x_*]$.
    \item We want to use a staggered grid scheme. Any given number $N \in \N$ with $N \cdot x_* \geq t_*$ determines the time step size
          \[
            \Delta t = \frac{t_*}{2N}.
          \]
          The time steps are
          \[
            t_{n} = (n-1) \Delta t, \quad n=1,\ldots,2N+1.
          \]
    \item Put
          \[
            M = \left \lfloor \frac{x_*}{t_*}\,N \right \rfloor \geq 1,
          \]
          then the spatial mesh size is
          \[
            \Delta x=\frac{x_*}{M},
          \]
          with the spatial grid points
          \[
            x_{j} = (j-1) \Delta x\,, \quad j=1,\ldots,N+M+1.
          \]
          Note that our scheme uses a trapezoidal computational domain $\mathcal{D}$ defined below that includes the target domain $[0,t_*] \times [0,x_*]$.
          Thereby, we can use all initial data that influence the solution on the target domain. In this way we avoid using a numerical boundary condition at $x_*$.
    \item The number
          \[
            \lambda=\frac{\Delta x}{2 \Delta t} \geq 1
          \]
          is used to satisfy the CFL-condition and to define the computational domain $\mathcal{D} = \left\{(t,x)\in \R^2\,:\,0 \leq t \leq t_*,\quad 0 \leq x \leq x_*+\lambda(t_*-t)\right \}.$
\end{enumerate}
The typical trapezoidal form of the computational domain is illustrated in Figure \ref{fig_domain}.
For the formulation and the stability of our numerical scheme we need two lemmas, given in \cite[Lemma 4.1]{KLW} and \cite[Lemma 4.2]{KLW}.
\begin{lem}\label{hilfssatz1}
    %
    Assume that $|b_{\pm}|< a_{\pm}$ and put $c_{\pm}=c(a_{\pm},b_{\pm})$ according to \eqref{cdefinition}. We recall that $\lambda\geq 1$.
    Then
    \begin{itemize}
        \item[a)] $-\left(a_- + \dfrac{b_-}{\lambda} \right) < b_- + \dfrac{c_-}{\lambda} < a_- + \dfrac{b_-}{\lambda},$
        \item[b)] $-\left(a_+ - \dfrac{b_+}{\lambda} \right) <  b_+ - \dfrac{c_+}{\lambda} < a_+ - \dfrac{b_+}{\lambda}.$
    \end{itemize}
\end{lem}
\begin{lem}\label{hilfssatz2}
    Assume that $a>0$, $0 < \eta \leq 1/3$ and $-a(1+\eta)<\xi<a(1-\eta)$. Then we obtain $4a^2(1+3\eta^2)-3\xi^2>0$ and
    \[
      \left|\frac{\xi + \eta\sqrt{4a^2(1+3\eta^2)-3\xi^2}}{1+3\eta^2}\right| < a.
    \]
\end{lem}
For the numerical discretization of the integral balance laws \eqref{weak_contour} we choose the triangular balance domain $\Omega$ depicted in Figure \ref{figeul}.
We assume that the midpoints $P_-=(\overline{t}, \overline{x}-\Delta x/2)$, $P_+=(\overline{t}, \overline{x}+\Delta x/2)$
and $P'=(\overline{t}+\Delta t, \overline{x})$ of the cords of $\partial \Omega$ are numerical grid points for the computational domain $\mathcal{D}$.
Let the numerical solution $(a_\pm,b_\pm)$ be given at the grid points $P_{\pm}$.
We have to require $|b_{\pm}|< a_{\pm}$ for the numerical solution in the actual time step $\overline{t}=t_n$ with $n=1,\ldots,2N$.
The major task is to calculate the numerical solution $(a',b')$ for the next time step $\overline{t}+\Delta t= t_{n+1}$ at its grid point $P'$, see Figure \ref{figeul}.
The spatial value $\overline{x} \geq 0$ is given. We have to determine a function
\begin{equation}\label{euler_solve}
    \text{Euler}(a_-,b_-,a_+,b_+,\overline{x},\Delta x, \lambda)=(a',b')
\end{equation}
for the calculation of $(a',b')$. This leads to the structure of a staggered grid scheme.
Note that at the boundary the balance region $\Omega$ may have parts outside $\mathcal{D}$, e.g.~points below the half-space $x\geq 0$.
In the latter case we employ a simple reflection principle for the numerical solution in order to use the function $\text{Euler}$ as well for the boundary points with $\overline{x}=0$.\\
Next we make use of the fact that the points $P_{\pm}$ with the numerical values $(a_{\pm},b_{\pm})$ and $P'$ with the unknown value $(a',b')$ are the 
\textit{midpoints of the three boundary cords} of the balance region $\Omega$.
We put $c_{\pm}=c(a_{\pm},b_{\pm})$ and $c'=c(a',b')$ for abbreviation, see \eqref{cdefinition}.
Then we use for $k=0,1,2$ the straight line paths $\gamma_k$ from Figure \ref{figeul}. For the corresponding path integrals
\[
  \int\limits_{\gamma_k}x a(t,x)\,dx-x b(t,x)\,dt\quad\text{and}\quad\int\limits_{\gamma_k}x b(t,x)\,dx-x c(a(t,x),b(t,x))\,dt
\]
with the unknown weak entropy solution $a(t,x)$, $b(t,x)$ the numerical discretizations $\tilde{I}_{k,a}$ and $\tilde{I}_{k,b}$, respectively, are given by
\begin{equation}\label{pathint0}
    \left\{
    \begin{aligned}
        %
        \tilde{I}_{0,a} &= \int\limits_{\gamma_0}x a'\,dx - x b'\,dt = a'\,\int \limits_{\overline{x} - \Delta x}^{\overline{x}+\Delta x}x\,dx = 4a'\lambda \Delta t \, \overline{x}\,,\\
        \tilde{I}_{0,b} &= \int\limits_{\gamma_0}x b'\,dx - x c'\,dt = b'\,\int \limits_{\overline{x} - \Delta x}^{\overline{x}+\Delta x}x\,dx = 4b'\lambda \Delta t \, \overline{x}\,.
        %
    \end{aligned}
    \right.
\end{equation}
\begin{equation}\label{pathint1}
    \left\{
    \begin{aligned}
        %
        \tilde{I}_{1,a} &= \int\limits_{\gamma_1}x a_+\,dx - x b_+\,dt = -2(\lambda a_+ - b_+) \Delta t\left( \overline{x} + \frac12 \Delta x\right)\,,\\
        \tilde{I}_{1,b} &= \int\limits_{\gamma_1}x b_+\,dx - x c_+\,dt = -2(\lambda b_+ - c_+) \Delta t\left( \overline{x} + \frac12 \Delta x\right)\,.
        %
    \end{aligned}
    \right.
\end{equation}
\begin{equation}\label{pathint2}
    \left\{
    \begin{aligned}
        %
        \tilde{I}_{2,a} &= \int\limits_{\gamma_2}x a_-\,dx - x b_-\,dt = -2(\lambda a_- + b_-) \Delta t\left( \overline{x} - \frac12 \Delta x\right)\,,\\
        \tilde{I}_{2,b} &= \int\limits_{\gamma_2}x b_-\,dx - x c_-\,dt = -2(\lambda b_- +c_-) \Delta t\left( \overline{x} -\frac12 \Delta x\right)\,.
        %
    \end{aligned}
    \right.
\end{equation}
We recall that $\overline{x}\geq 0$ and distinguish two cases.\\
\noindent
\underline{\textbf{Case 1:}} First we assume that $\overline{x} > 0$. In this case we put
\begin{equation}\label{qdef}
    \tilde{q} = \frac{\Delta x}{2\overline{x}} \leq 1.
\end{equation}
The numerical discretization of the first balance law in \eqref{weak_contour} gives
\begin{equation}\label{abilanz}
    \tilde{I}_{0,a}=-\tilde{I}_{1,a}-\tilde{I}_{2,a}\,.
\end{equation}
We obtain from \eqref{pathint0}, \eqref{pathint1},\eqref{pathint2}, \eqref{qdef} and \eqref{abilanz} for $a'$ the explicit solution
\begin{equation}\label{aC}
    a' = \frac12\left(a_- +\frac{b_-}{\lambda}\right)(1-\tilde{q}) + \frac12\left(a_+ - \frac{b_+}{\lambda}\right)(1+\tilde{q}).
\end{equation}
For the numerical discretization of the second balance law in \eqref{weak_contour} we approximate the integral
\begin{equation*}
    \frac12 \iint \limits_{\Omega}(a-c)\,dtdx\quad\text{by}\quad \frac12 (a'- c')\iint \limits_{\Omega} dtdx = (a'- c')\Delta t \Delta x.
\end{equation*}
Now \eqref{pathint0}, \eqref{pathint1}, \eqref{pathint2} give the following ansatz for the calculation of $b'$:
\begin{equation}\label{bbilanz}
    \begin{split}
        \tilde{I}_{0,b} &= -\tilde{I}_{1,b}-\tilde{I}_{2,b}+(a'- c')\Delta t \Delta x,\\
        \Leftrightarrow\quad
        b' &= \frac{1}{2}\left(b_+ - \frac{c_+}{\lambda}\right)(1 + \tilde{q}) + \frac{1}{2}\left(b_- + \frac{c_-}{\lambda}\right)(1 - \tilde{q}) + (a' - c')\frac{\tilde{q}}{2\lambda}.
    \end{split}
\end{equation}
Recall that $c = c(a,b)$ is given by \eqref{cdefinition} and thus the values $c_\pm$ are known. We further use \eqref{cdefinition} to substitute $c'$ and obtain
\begin{equation*}
    b' - \frac{\tilde{q}}{3\lambda}\sqrt{4a'^2 - 3b'^2} = \frac{1}{2}\left(b_+ - \frac{c_+}{\lambda}\right)(1 + \tilde{q}) + \frac{1}{2}\left(b_- + \frac{c_-}{\lambda}\right)(1 - \tilde{q})
    - a'\frac{\tilde{q}}{3\lambda}.
\end{equation*}
This is an implicit equation for $b'$ where the right-hand side is known.
Introducing the abbreviations
\begin{equation}\label{Cxieta}
    \eta = \frac{\tilde{q}}{3\lambda}\;\text{and}\;\xi = \frac12\left(b_- +\frac{c_-}{\lambda}\right)(1-\tilde{q}) + \frac12\left(b_+ - \frac{c_+}{\lambda}\right)(1+\tilde{q})-a' \eta
\end{equation}
we obtain the implicit equation
\begin{equation}\label{bCimp}
    b' = \xi+\eta \sqrt{4a'^2-3b'^2}.
\end{equation}
This leads to a quadratic equation for $b'$. Lemma \ref{hilfssatz1} gives
\[
  -a'(1+\eta) < \xi < a'(1-\eta)
\]
for the quantity $a'$ in \eqref{aC}. In order to apply Lemma \ref{hilfssatz2} with $a'$ instead of $a$ we have to choose the solution
\begin{equation}\label{bCexplicit}
    b' = \frac{\xi + \eta\sqrt{4a'^2(1+3\eta^2)-3\xi^2}}{1+3\eta^2}
\end{equation}
of \eqref{bCimp} with the positive square root. Now $b'$ is well defined with $|b'|<a'$, see the transformation \eqref{statetrans} in state space.\\
\noindent
\underline{\textbf{Case 2:}} We assume that $\overline{x}=0$. In this case we put
\begin{equation}\label{case2}
    a' = a_+-b_+/\lambda\,, \quad b' = 0.
\end{equation}
Here we apply the reflection method from \cite[Remark 4.5, equation (4.12)]{KLW}. We note that $|b_+|<a_+$ implies $a' >0$ and hence $|b'|=0<a'$ also in case 2.
We summarize our results in the following
\begin{thm}[Numerical solution $(a',b')$ for the balance region $\Omega$]\label{apbp} 
    Given are real quantities $\overline{x}\geq 0$ and $a_{\pm}$, $b_{\pm}$. Assume that $|b_{\pm}|< a_{\pm}$.
    We recall $\lambda\geq 1$ defined in terms of $\Delta t$ and $\Delta x$ and put $c_{\pm}=c(a_{\pm},b_{\pm})$ 
    in  \eqref{cdefinition}.
    \begin{enumerate}[(i)]
        \item For $\overline{x}>0$ we calculate  $a'$ and $b'$ from \eqref{qdef}, \eqref{aC}, \eqref{Cxieta} and \eqref{bCexplicit}.
        \item For $\overline{x}=0$ we calculate  $a'$ and $b'$ from \eqref{case2}.
    \end{enumerate}
    Then we have $|b'|<a'$ in both cases.
    \dokendDef
\end{thm}
\begin{defin}[The function Euler]\label{eulerfunction}
    The state $(a',b')$ from Theorem \ref{apbp} defines the function $\text{Euler}$ in \eqref{euler_solve}.
    \dokendDef
\end{defin}
Now we are able to formulate the numerical scheme for the solution of the initial-boundary value problem \eqref{initial_final}, \eqref{weak_contour}.
We  construct staggered grid points in the computational domain $\mathcal{D}$ and compute the numerical solution at these grid points.
Using the function $\text{Euler}$ we obtain the evolution of the numerical solution in time,
i.e., it allows us to construct the solution at time $t=t_{n+1}$ from the solution which is already calculated in the grid points   at the former time step $t=t_n$.
\begin{enumerate}[(I)]
    \item The staggered grid points are $(t_n,x_{n,j}) \in \mathcal{D}$ for $t_n=(n-1)\Delta t$, $n=1,\ldots,2N+1$ and $j=1,\ldots,M+N-\lfloor (n-1)/2 \rfloor$ with
          \begin{align*}
            x_{n,j} = 
            \begin{cases}
                (x_{j}+x_{j+1})/2\; &\text{if $n$ is odd}\\
                x_j\; &\text{if $n$ is even}.
            \end{cases}
          \end{align*}
          We want to calculate the numerical solution $(a_{n,j},b_{n,j})$ at $(t_n,x_{n,j})$.
    \item For $j = 1,\ldots,M+N$ we calculate the numerical solution $(a_{1,j}, b_{1,j})$ at the grid point $(t_1,x_{1,j})=(0,(x_j+x_{j+1})/2)$ from the given initial data by
          \[
            a_{1,j} = a_0\left(x_{1,j}\right),\quad b_{1,j} = b_0\left(x_{1,j}\right).
          \]
          This corresponds to taking the integral average of the initial data on $(x_j,x_{j+1})$ and using the midpoint rule as quadrature.
    \item Assume that for a fixed \textit{odd index} $n\in \{1,\ldots,2N\}$ we have already determined the numerical solution $(a_{n,j}, b_{n,j})$
          at the grid points $(t_n,x_{n,j})$, $j=1,\ldots,M + N - (n - 1)/2$.\\
          First we determine the solution $(a_{n+1,1},b_{n+1,1})$ at the boundary point $(t_{n+1},x_{n+1,1}) = (t_{n+1},0)$ according to \eqref{case2}.
          For this purpose we put $a_+ = a_{n,1}$, $b_+ = b_{n,1}$, $a_- = a_{n,1}$, $b_- = -b_{n,1}$ and have
          \[
            (a_{n+1,1}, b_{n+1,1}) = \text{Euler}(a_-,b_-,a_+,b_+,0,\Delta x,\lambda)\;\text{with}\; b_{n+1,1} = 0.
          \]
          Next we put $a_- = a_{n,j-1}$, $b_- = b_{n,j-1}$ and $a_+ = a_{n,j}$, $b_+ = b_{n,j}$ for $j = 2,\ldots,M + N  - (n - 1)/2$ and determine the values $a_{n+1,j}$, $b_{n+1,j}$ 
          at time $t_{n+1}$ and position $\overline{x} = x_{n+1,j} = x_{j}$ from
          \[
            (a_{n+1,j}, b_{n+1,j})=\text{Euler}(a_-,b_-,a_+,b_+,\overline{x},\Delta x,\lambda).
          \]
    \item Assume that for a fixed \textit{even index} $n \in \{1,\ldots,2N\}$ we have already determined the numerical solution $(a_{n,j}, b_{n,j})$
    at the grid points $(t_n,x_{n,j})$, $j=1,\ldots,M + N - n/2 + 1$.\\
    We put $a_- = a_{n,j}$, $b_- = b_{n,j}$ and $a_+ = a_{n,j+1}$, $b_+ = b_{n,j+1}$ for $j = 1,\ldots,M + N - n/2$ and determine the values $a_{n+1,j}$, $ b_{n+1,j}$ at time $t_{n+1}$ and position 
    $\overline{x} = x_{n+1,j} = (x_{j} + x_{j+1})/2$ from
    \[
      (a_{n+1,j}, b_{n+1,j}) = \text{Euler}(a_-,b_-,a_+,b_+,\overline{x},\Delta x,\lambda).
    \]
\end{enumerate}
Based on Lemma \ref{hilfssatz1} and \ref{hilfssatz2} we obtain Theorem \ref{apbp}. 
Using \eqref{initial_final} we can state the following
\begin{thm}\label{stable_scheme} 
    The numerical scheme described above preserves a positive pressure $p>0$, provided the pressure is positive in the given initial data.
    \dokendDef
\end{thm}
%
\section{Numerical Results}\label{sec:numeric}
%
We perform extensive numerical computations where we first validate the one-dimensional radially symmetric   scheme (RadSymS) presented in Sect.~\ref{scheme1} by means of self-similar solutions. Then we verify that the one-dimensional scheme can be used for the validation of genuinely multi-dimensional solvers.
For this purpose we briefly summarize the numerical methods and models in Sect.~\ref{sec:num_meth}.
Then we set up different configurations for which we perform the computations in Sect.~\ref{configurations}.

Finally, we compare the numerical results in Sect.~\ref{sec:discussion}.
%
\subsection{Numerical  methods}\label{sec:num_meth}
~\newline
\textbf{\emph{One-dimensional radially symmetric solver RadSymS.}}
As described in Sect.~\ref{intro} a radially symmetric solution of the ultra-relativistic Euler equations satisfies the quasi one-dimensional problem
%
\begin{equation}\label{radsys}
    \left\{
    \begin{aligned}
        %
        \quad\quad\quad\frac{\partial}{\partial t}\left(x^{d-1}a \right) + \frac{\partial}{\partial x}\left(x^{d-1}b \right) &= 0,\\
        \quad\quad\quad\frac{\partial}{\partial t}\left(x^{d-1}b \right) + \frac{\partial}{\partial x}\left(x^{d-1}c \right) &= \frac{d-1}{2} x^{d-2} (a - c),\\
        \lim \limits_{t \searrow 0}a(t,x) = a_0(x),\quad \lim \limits_{t \searrow 0}b(t,x) &= b_0(x)\,.
    \end{aligned}\right.
    %
\end{equation}
for $t>0$ and $x > 0$ representing the radial direction.
Here $d=2$ or $d=3$ denotes the space dimension and $c = c(a,b)$ is given by \eqref{cdefinition}.
We are looking for weak solutions $a = a(t,x)$, $b = b(t,x)$ with $|b| < a$.
For $d=2$ the weak formulation of 
the system \eqref{radsys}
is given in \eqref{weak_contour}, and for $d=3$ in \cite[Equation (2.13)]{KLW}.
Using the transformation \eqref{statetrans}, its inverse \eqref{inverse_statetrans} and the velocity
\[
  v = \frac{u}{\sqrt{1+u^2}}\quad\text{with}\;|v| < 1
\]
we replace the state variables $a$ and $b$ by $p$ and $v$, respectively.
We prescribe the initial pressure $p_0=p(0,\cdot)$ as well as the initial velocity $v_0=v(0,\cdot)$.
Here we have chosen the variable $v$ because the restriction $|v| < 1$ leads to better color plots (the variable $u$ is unbounded).

The quasi one-dimensional problem \eqref{radsys} is approximately solved by the one-dimensional radially symmetric scheme presented  in Sect.~\ref{scheme1} with $N=5000$.
Note that for constant pressure $p_0 > 0$  and $v_0 = 0$, we obtain a stationary solution.
This solution is exactly reconstructed by the one-dimensional radially symmetric scheme in Section \ref{scheme1}.
Such a steady part is contained in the solutions to the examples presented below.

\textbf{\emph{Description of the multi-dimensional DG solver MultiWave.}}
We also compare the results with the numerical solution of the original multi-dimensional initial value problem for the ultra-relativistic Euler equations \eqref{energy_general},\eqref{momentum_general}.
For $j=1,\dots,d$ it reads
\begin{equation}\label{orisys}
    \left\{
    \begin{aligned}
        %
        \frac{\partial}{\partial t}\left(3\tilde{p} + 4\tilde{p}|\mathbf{u}|^2 \right)
        + \sum\limits_{k=1}^d\frac{\partial}{\partial x_k}\left(4\tilde{p}u_k\sqrt{1 + |\mathbf{u}|^2} \right) &= 0,\\
        \frac{\partial}{\partial t}\left(4\tilde{p}u_j \sqrt{1 + |\mathbf{u}|^2} \right)
        + \sum\limits_{k=1}^d\frac{\partial}{\partial x_k}\left(\tilde{p}\delta_{jk} + 4\tilde{p}u_j u_k\right) &= 0,\\
        \tilde{p}(0,\mathbf{x}) = \tilde{p}_0(|\mathbf{x}|), \quad \mathbf{u}(0,\mathbf{x}) = \mathbf{u}_0(\mathbf{x}) &= u_0(|\mathbf{x}|) \frac{\mathbf{x}}{|\mathbf{x}|}.
    \end{aligned}\right.
    %
\end{equation}
We require radially symmetric solutions of this system, i.e., for $t \geq 0$ and $x=|\mathbf{x}|>0$ the restrictions for pressure and velocity are given by
\[
  \tilde{p}(t,\mathbf{x})=p(t,x)\,,\quad \mathbf{u}(t,\mathbf{x})=\frac{v(t,x)}{\sqrt{1 - v(t,x)^2}} \cdot \frac{\mathbf{x}}{x}\,.
\]
For $t=0$ we obtain given initial data $\tilde{p}(0,\mathbf{x}) =p(0,x)=p_0(x)$ and
\[
  \mathbf{u}(0,\mathbf{x}) = \mathbf{u}_0(\mathbf{x}) = \frac{v_0(x)}{\sqrt{1 - v_0(x)^2}} \cdot \frac{\mathbf{x}}{x}\,,
\]
and for $\mathbf{x}=\mathbf{0} \in \R^d$ we may also put $\mathbf{u}(0,\mathbf{0})=\mathbf{0}$ and $v_0(0)=0$.

The ultra-relativistic  Euler equations are solved
using a classical Runge-Kutta discontinuous Galerkin (RK-DG) method \cite{Cockburn:1998jt}.
%
%
For the implementation of this system we need the eigenvalues and corresponding left- and right-eigenvectors of the flux Jacobian. Details can be found in the Appendix \ref{app:eig_sys_ure}.\\
Here we apply a third order DG scheme using  piecewise polynomial elements of order $p=3$ 
and a third-order explicit SSP-Runge-Kutta method with three stages for the time discretization. 
For a numerical flux we choose the local Lax-Friedrichs flux. The Gibbs phenomena near to discontinuites is suppressed by the minmod limiter from \cite{Cockburn:1998jt}. Because of  the explicit time discretization we restrict the timestep size by means of a CFL number.
%
The efficiency of the scheme is improved by local grid adaption where we employ the
multiresolution concept based on multiwavelets.
The key idea is to perform a multiresolution analysis on a sequence of
nested grids providing a decomposition of the data on a coarse scale and a sequence
of details that encode the difference of approximations on
subsequent resolution levels. The detail coefficients become small when the underlying
data are locally smooth and, hence, can be discarded when dropping below a
threshold value $\varepsilon_{thresh}$.
By means of the thresholded sequence a new, locally refined grid
is determined. Details on this concept can be found in 
\cite{HovhannisyanMuellerSchaefer-2014,GerhardIaconoMayMueller-2015,GerhardMueller-2016,GerhardMuellerSikstel:2021}.\\
For all examples, we set the CFL number to $CFL=0.06$. Table~\ref{tab:parameter-multiwave} summarizes the computational domain $\Omega\subset\R^2$, the maximum refinement level $L\in\N$ and the cell size $\Delta\mathbf{x}_0 \in \R^2$ at the coarsest refinement level for each example. Due to the dyadic grid hierarchy in MultiWave the cell size at refinement level $l$ is $\Delta\mathbf{x}_l = 2^{-l}\Delta\mathbf{x}_0$, for $l=0,\dots,L$.

\begin{table}[h!]
    \centering
    \begin{tabular}{c|ccccc}
                    & Ex.~1   & Ex.~2  & Ex.~3 & Ex.~4 & Ex.~5     \\
        \hline
        $\Omega$     & $[-2,2]^2$ & $[-2,2]^2$    & $[-6,6]^2$   & $[-6,6]^2$  &  $[-5,5]^2$  \\
        $L$         & 8 &  8   & 9   & 9  &  8  \\
        $\Delta\mathbf{x}_0$   &  $(2,2)$   &  $(2,2)$   & $(4,4)$   & $(4,4)$  &  $(2,2)$    \\
    \end{tabular}
    %
    \caption{Discretization parameters for MultiWave.}
    \label{tab:parameter-multiwave}
\end{table}

%
%
\subsection{Benchmark Tests} \label{configurations}
In the following we  set up several radially symmetric problems where two of them provide self-similar solutions, see Example 1 and 2. All of these configurations may serve as benchmark problems for the validation of multi-dimensional solvers, e.g., finite volume schemes or discontinuous Galerkin (DG) schemes.

\textbf{\emph{Example 1: Solutions including a shock and a stationary part.}}
%
Following \cite{GL} self-similar solutions can be constructed solving an ODE system in radial direction $x$. These solutions are in particular constant along rays $\xi=x/t$ or, equivalently, $\vartheta=t/x$. Such solutions are used for validation purposes for both the one-dimensional radially symmetric solver and the multi-dimensional DG solver.\\
We consider constant initial data with pressure $p_0=1$ and radial velocity $v_0 \in (-1,0)$.
Due to \cite[Section 2.3]{GL} there is a solution $p(t,x)=P(\vartheta)$ and $v(t,x)=V(\vartheta)$ depending only on $\vartheta=t/x$ for $t,x >0$,
with a single straight line shock emanating from the zero point. Let $\tilde{s} \in (0,1)$ be the unknown constant shock speed.
Then we put $v_-=0$ and can find an unknown pressure $p_->0$ with
\[
  p(t,x)=p_-\,, \quad v(t,x)=v_-=0 \quad \mbox{for~} x < \tilde{s}\cdot t.
\]
Due to the Rankine-Hugoniot shock conditions introduced in \cite[Section 4.4]{Kunikthesis} we obtain from \cite[page 82]{Kunikthesis}
for a so called 3-shock after a lengthy calculation the algebraic shock conditions
\begin{equation*}\label{pplus}
  \frac{p_+}{p_-}=\frac{9\tilde{s}^2-1}{3(1-\tilde{s}^2)},\quad v_+=\frac{3}{2}\tilde{s}-\frac{1}{2\tilde{s}}\,, \quad v_-=0,
\end{equation*}
\begin{equation*}\label{ungl}
  0 < p_+ < p_-,\quad \frac13 < \tilde{s} < \frac{1}{\sqrt{3}}.
\end{equation*}
Here $p_-,v_-$ are the values of pressure and velocity left to the 3-shock, and $p_+,v_+$ are the values of pressure and velocity right to the 3-shock.
Due to Lai \cite[Section 2.3]{GL} the solution $p=P(\vartheta)$, $v=V(\vartheta)$ satisfies the initial value problem of ordinary differential equations
\begin{equation}\label{GL_ODE_sys}
   \begin{split}
   \dot{V}(\vartheta) &= (d-1)\frac{V(V-\vartheta)(1-V^2)}{3(\vartheta V-1)^2-(V-\vartheta)^2},\quad V(0)=v_0 \in (-1,0),\\
   \dot{P}({\vartheta}) &= (d-1)\frac{4PV(\vartheta V-1)}{3(\vartheta V-1)^2-(V-\vartheta)^2},\quad P(0)=p_0=1.
   \end{split}
\end{equation}
Moreover, we show in Appendix \ref{app:lai} that Lai's results guarantee a unique solution for $0 \leq \vartheta < \vartheta_{max}$ 
with a certain value $\vartheta_{max} >\sqrt{3}$ and a unique value $\tilde{\vartheta} \in (\sqrt{3},\min(3,\vartheta_{max}))$ such that
\begin{equation}\label{vvalue}
   V(\tilde{\vartheta})=\frac{3}{2\tilde{\vartheta}}-\frac{\tilde{\vartheta}}{2}.
\end{equation}
After 
having determined $\tilde{\vartheta}$ we finally obtain
\begin{equation}\label{shockvalues}
   \tilde{s}=\frac{1}{\tilde{\vartheta}},\quad v_+=V(\tilde{\vartheta}),\quad p_+=P(\tilde{\vartheta}),\quad p_-=p_+\frac{3(1-\tilde{s}^2)}{9\tilde{s}^2-1}.
\end{equation}
For our numerical simulation we choose $d=2$,
a constant initial pressure $p_0 = 1$ and a constant initial velocity $v_0=-1/\sqrt{2}$.
This corresponds to the following initial data for the original initial value problem \eqref{orisys}
\[
  \tilde{p}(0,\mathbf{x}) = 1\quad\text{and}\quad\mathbf{u}(0,\mathbf{x}) = -\frac{\mathbf{x}}{|\mathbf{x}|}\,, \quad\mathbf{x}\in\R^d \setminus \{\mathbf{0} \}.
\]
The radially symmetric solution of \eqref{orisys} is determined by the solution of the ODE system \eqref{GL_ODE_sys} which gives a shock wave moving at constant speed $\tilde{s}$. We want to point out that these ODEs are written in terms of $\vartheta = t/x$. Thus the initial data $P(0)$ and $V(0)$ for \eqref{GL_ODE_sys} prescribe the solution for \eqref{orisys} at infinity given by $p_0 = P(0)$ and $v_0 = V(0)$.
The ODE system \eqref{GL_ODE_sys} is solved by applying the classical fourth order RK-scheme with step size $h = 10^{-6}$.
For the computation of   $\tilde{\vartheta}$ the ODE solver is run until \eqref{vvalue}
 is satisfied with a tolerance of $\varepsilon = 10^{-9}$.
%
The solution values for the shock speed, the states ahead and behind of the shock, respectively, are presented in Table \ref{tab:shock_dat_ex1}.
\begin{table}[h!]
    \centering
    \begin{tabular}{c|ccccc}
                    & $\tilde{s}$   & $p_-$         & $v_-$ & $p_+$         & $v_+$     \\
        \hline
        $d = 2$     & $0.45503$     & $15.75505$    & $0$   & $\hphantom{1}5.71869$    & $-0.41629$ \\
        $d = 3$     & $0.52314$     & $25.56463$    & $0$   & $17.16524$    & $-0.17106$
    \end{tabular}
    %
    \caption{Example 1: Shock states for $d=2,3$}
    \label{tab:shock_dat_ex1}
\end{table}
Here the numerical values for $\tilde{s}$ and for $p_-, v_-$ on the plateau are the same up to three digits after the decimal point.
We want to mention
that for $d=3$ the values given in \eqref{vvalue} and \eqref{shockvalues} 
show a good agreement with the numerical results given in 
\cite[Section 5, Example 3]{KLW}.

\textbf{\emph{Example 2:  Self-Similar Expansion.}\label{selfsim_expansion}}
In this example we prescribe initial data leading to a smooth self-similar solution
by applying Lai's approach as in Example 1.
Note that for $d=3$ the solution will expand into vacuum at the speed of light when the initial value for $v_0$ is close enough to one.
We do not consider this case here and refer to \cite{GL} for more details.
For our numerical simulation we choose $d=2$, a constant initial pressure $p_0 = 1$ and a constant initial velocity $v_0 = 1/\sqrt{2}$.
This corresponds to the following initial data for the original initial value problem \eqref{orisys}
\[
  \tilde{p}(0,\mathbf{x}) = 1\quad\text{and}\quad\mathbf{u}(0,\mathbf{x}) = \frac{\mathbf{x}}{|\mathbf{x}|}\,, \quad\mathbf{x}\in\R^d \setminus \{\mathbf{0} \}.
\]

The solution consists of a rarefaction wave that is determined by the ODE \eqref{GL_ODE_sys}. In contrast to Example 1 there is no shock.
\textbf{\emph{Example  3: Expansion of a Spherical Bubble.}}
Next we consider the expansion of a spherical bubble in $d=2$ with the following initial data
\begin{equation*}
  p_0(x) = \begin{cases}
      1 \quad & \mbox{for~} 0 \leq x \leq 1\\
      0.1 \quad & \mbox{for~} x > 1,
  \end{cases}
  \qquad v_0(x) = 0.
\end{equation*}
These initial values correspond to the following initial values for the original initial value problem \eqref{orisys}
\begin{equation*}
  \tilde{p}(0,\mathbf{x}) = \begin{cases}
      1 \quad & \mbox{for~} 0 \leq |\mathbf{x}| \leq 1\\
      0.1 \quad & \mbox{for~} |\mathbf{x}| > 1,
  \end{cases}
  \qquad \mathbf{u}(0,\mathbf{x}) = \mathbf{0}.
\end{equation*}
\textbf{\emph{Example  4:  Collapse of a Spherical Bubble.}}
Next we study the collapse of a spherical bubble in 
$d=2$
with the following initial data
\begin{equation*}
  p_0(x) = \begin{cases}
      0.1 \quad & \mbox{for~} 0 \leq x \leq 1\\
      1 \quad & \mbox{for~} x > 1,
  \end{cases}
  \qquad v_0(x) = 0.
\end{equation*}
These initial values correspond to the following initial values for the original initial value problem \eqref{orisys}
\begin{equation*}
  \tilde{p}(0,\mathbf{x}) = \begin{cases}
      0.1 \quad & \mbox{for~} 0 \leq |\mathbf{x}| \leq 1\\
      1 \quad & \mbox{for~} |\mathbf{x}| > 1,
  \end{cases}
  \qquad \mathbf{u}(0,\mathbf{x}) = \mathbf{0}.
\end{equation*}
\textbf{\emph{Example 5: Initially Periodic Radial Velocity.}}
Finally, we study initial data with a periodic velocity in radial direction, i.e.,
\begin{align*}
    p_0(x) = 1,\quad u_0(x) = \begin{cases}
        \sin(2\pi x),\;&x < 1,\\
        0,\;&x \geq 1.
    \end{cases}
\end{align*}
These correspond to the following initial data for the multi-dimensional simulation
\begin{align*}
    \tilde{p}(0,\mathbf{x}) = 1,\quad \mathbf{u}(0,\mathbf{x}) = \begin{dcases}
        \sin(2\pi |\mathbf{x}|)\frac{\mathbf{x}}{|\mathbf{x}|},\;&|\mathbf{x}| < 1,\\
        0,\;&\mathbf{x} \geq 1.
    \end{dcases}
\end{align*}

\subsection{Discussion of results}\label{sec:discussion}
\begin{figure}[h!]
    %
    %
    %
    %
%
    %
    \subfigure[Results for the pressure $p$]{
    \includegraphics[width=\textwidth]{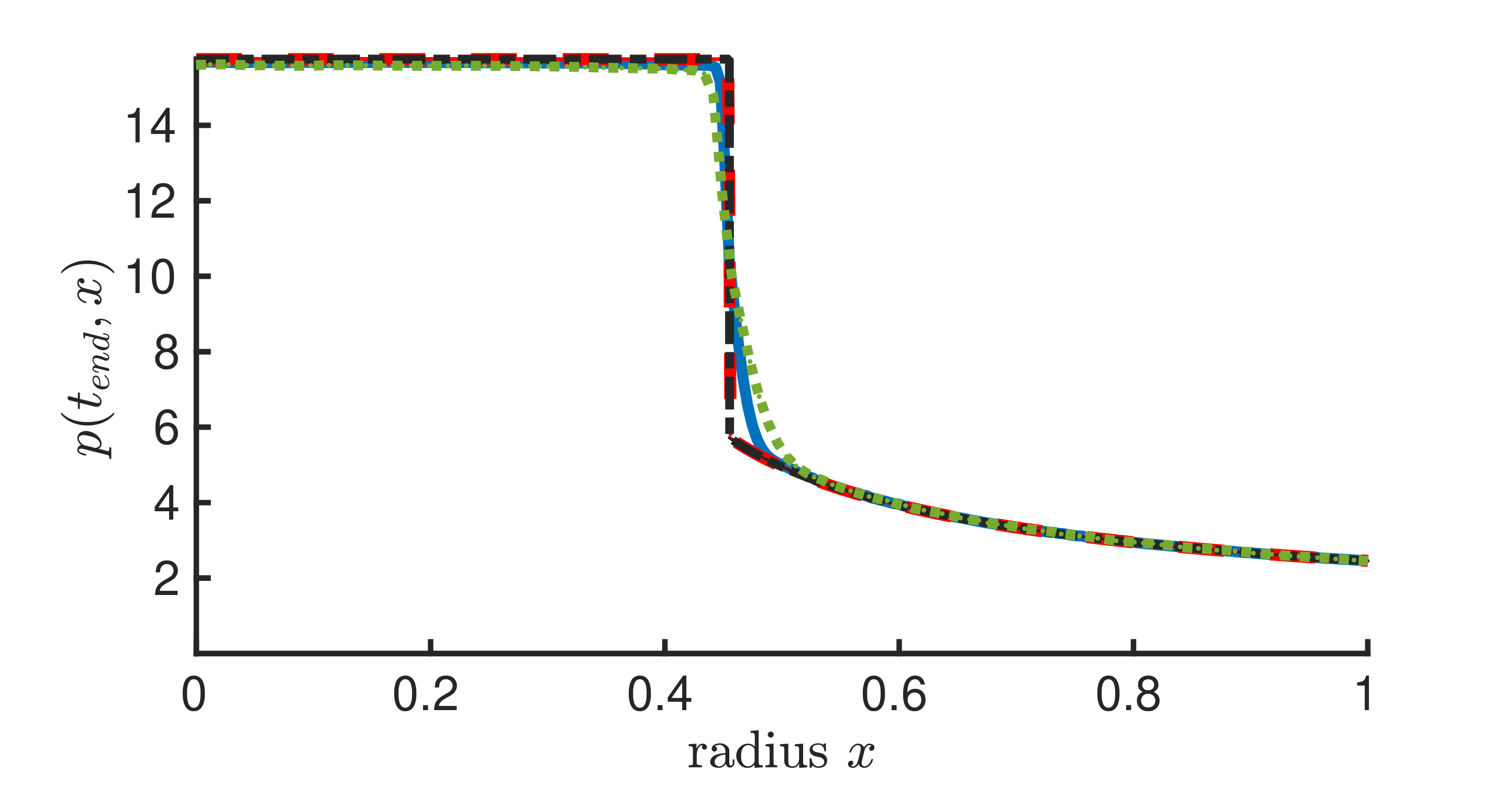}}
    \subfigure[Results for the velocity $v$]{
    \includegraphics[width=\textwidth]{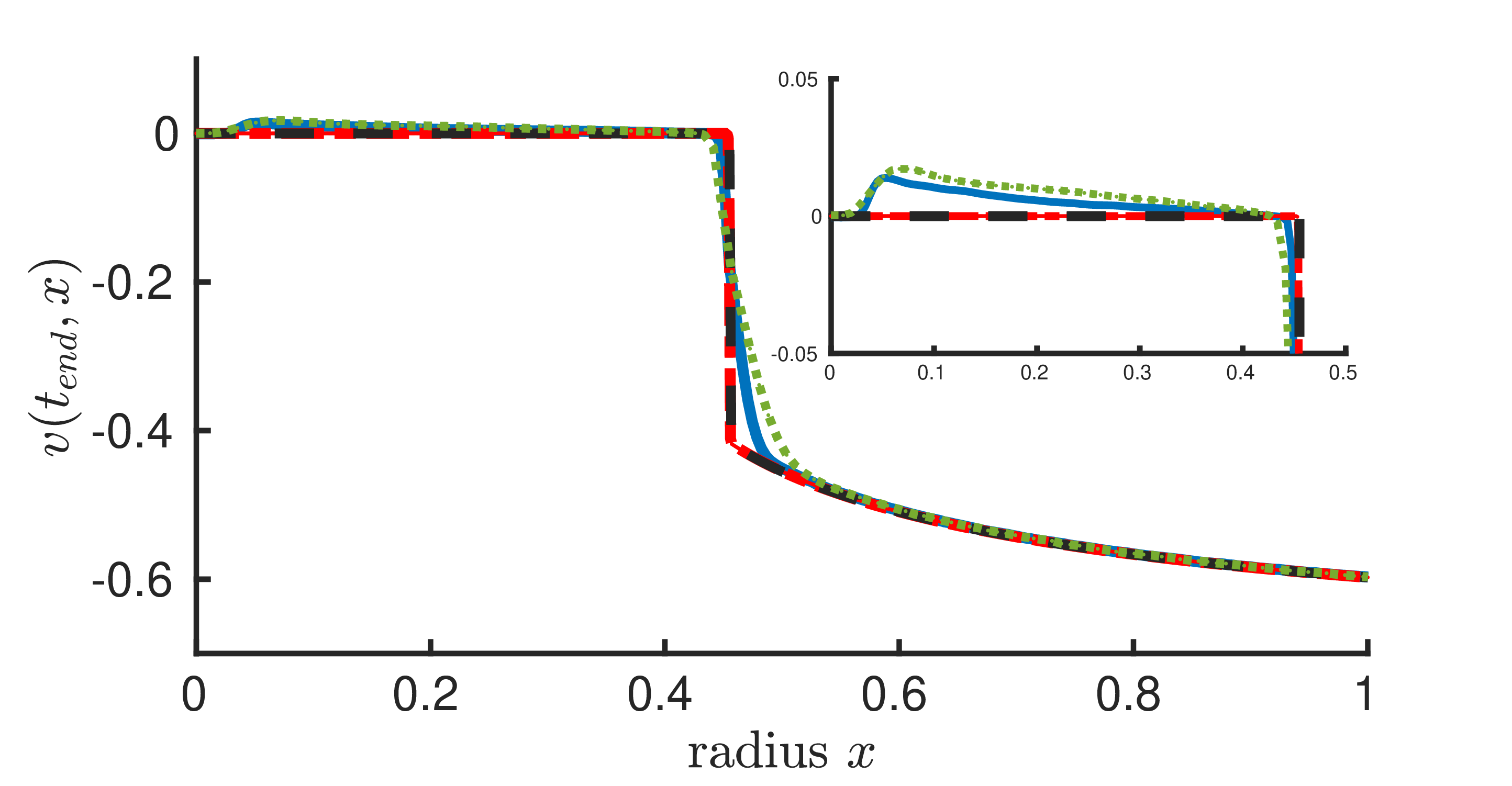}}
    \caption{   \textbf{Example 1:} 
     Comparison for MultiWave (green $L = 8$, blue $L = 9$) and RadSymS (red) in radial direction  with the solution of ODE system \eqref{GL_ODE_sys} (black) at final time $t_{end} = 1$ for the pressure $p$, see (a), and velocity $v$, see (b). 
  }
    \label{fig:example-1}
\end{figure}

\begin{figure}[h!]
    %
    %
    %
    %
%
    %
    \subfigure[Results for the pressure $p$]{
    \includegraphics[width=\textwidth]{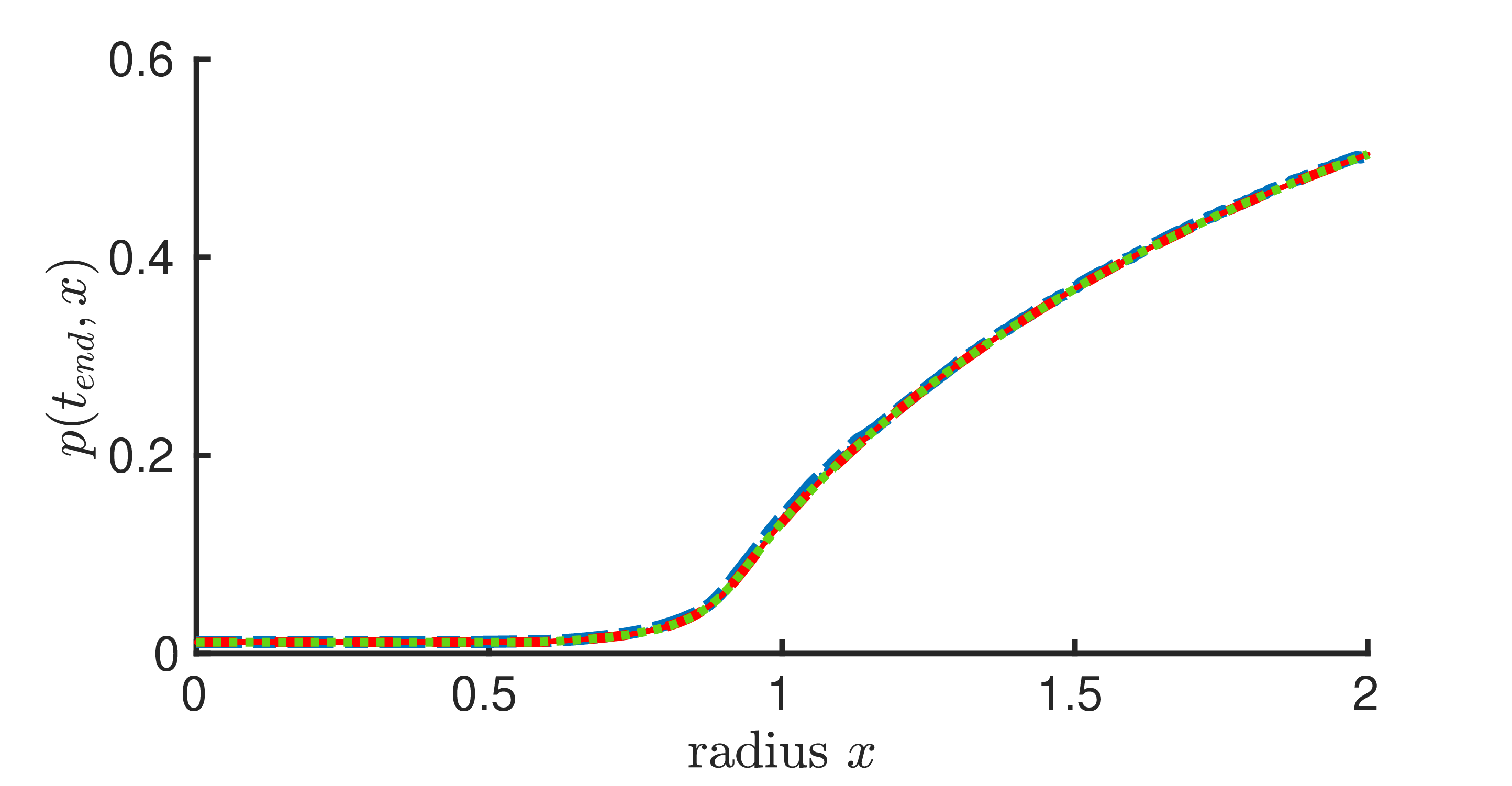}}
    \subfigure[Results for the velocity $v$]{
    \includegraphics[width=\textwidth]{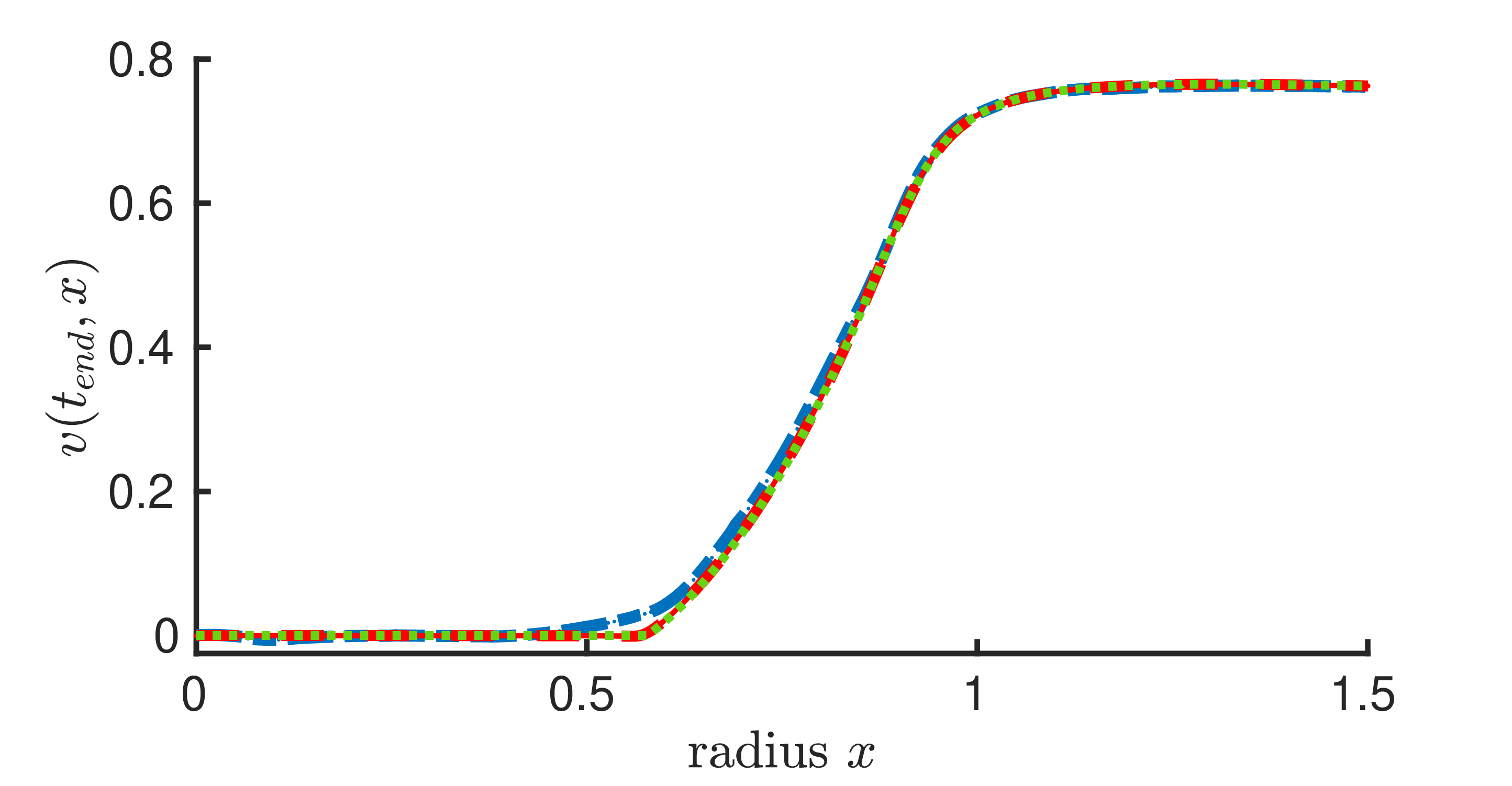}}
    %
    \caption{\textbf{Example 2:} 
     Comparison for MultiWave (blue) and RadSymS (red) in radial direction  with the solution of ODE system \eqref{GL_ODE_sys} (green) at final time $t_{end} = 1$ for the pressure $p$, see (a), and velocity $v$, see (b).}
    \label{fig:example-2}
\end{figure}

\begin{figure}[h!]
    \subfigure[Results for the pressure $p$]{
    \includegraphics[width=\textwidth]{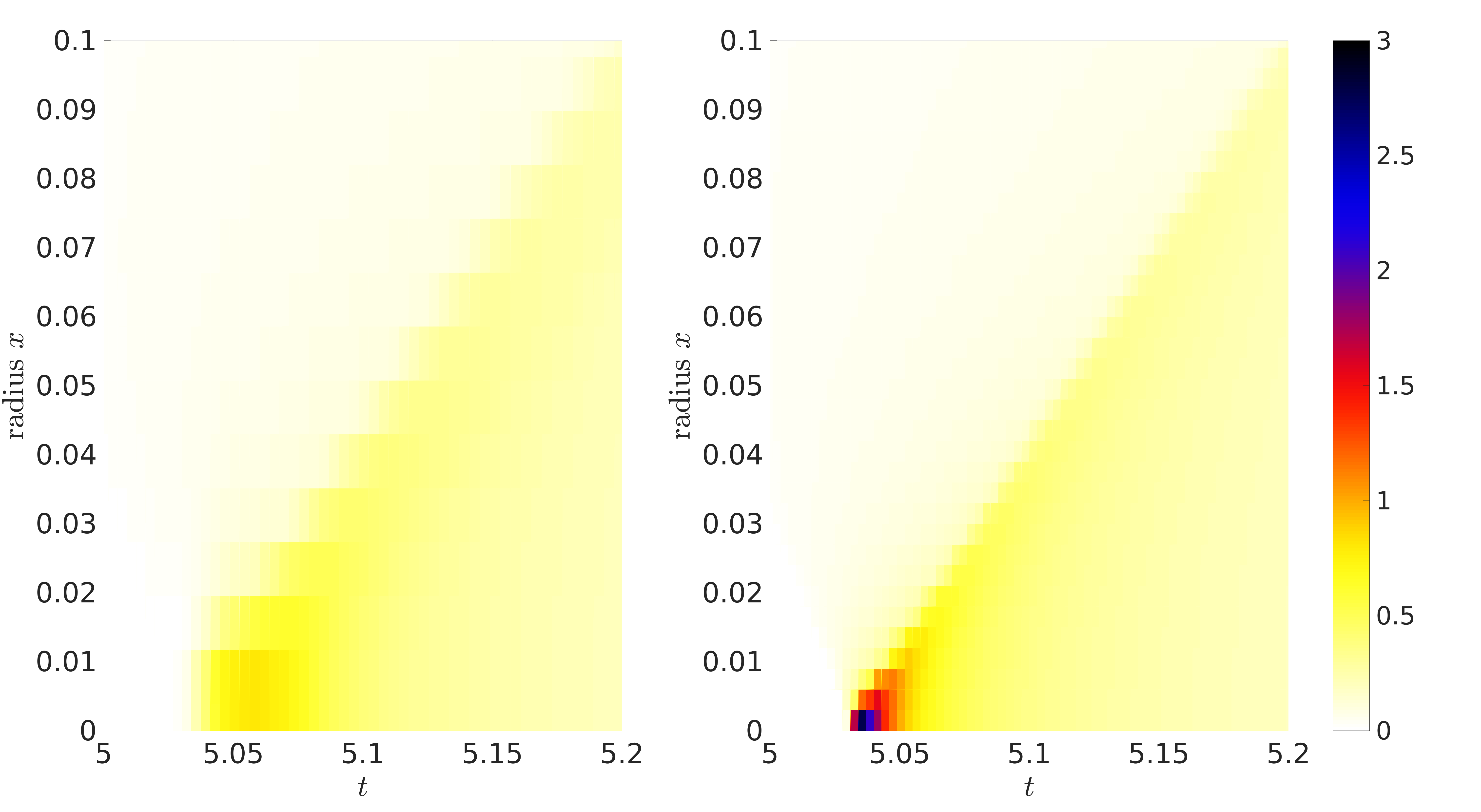}
    }\\
    \subfigure[Results for the velocity $v$]{
    \includegraphics[width=\textwidth]{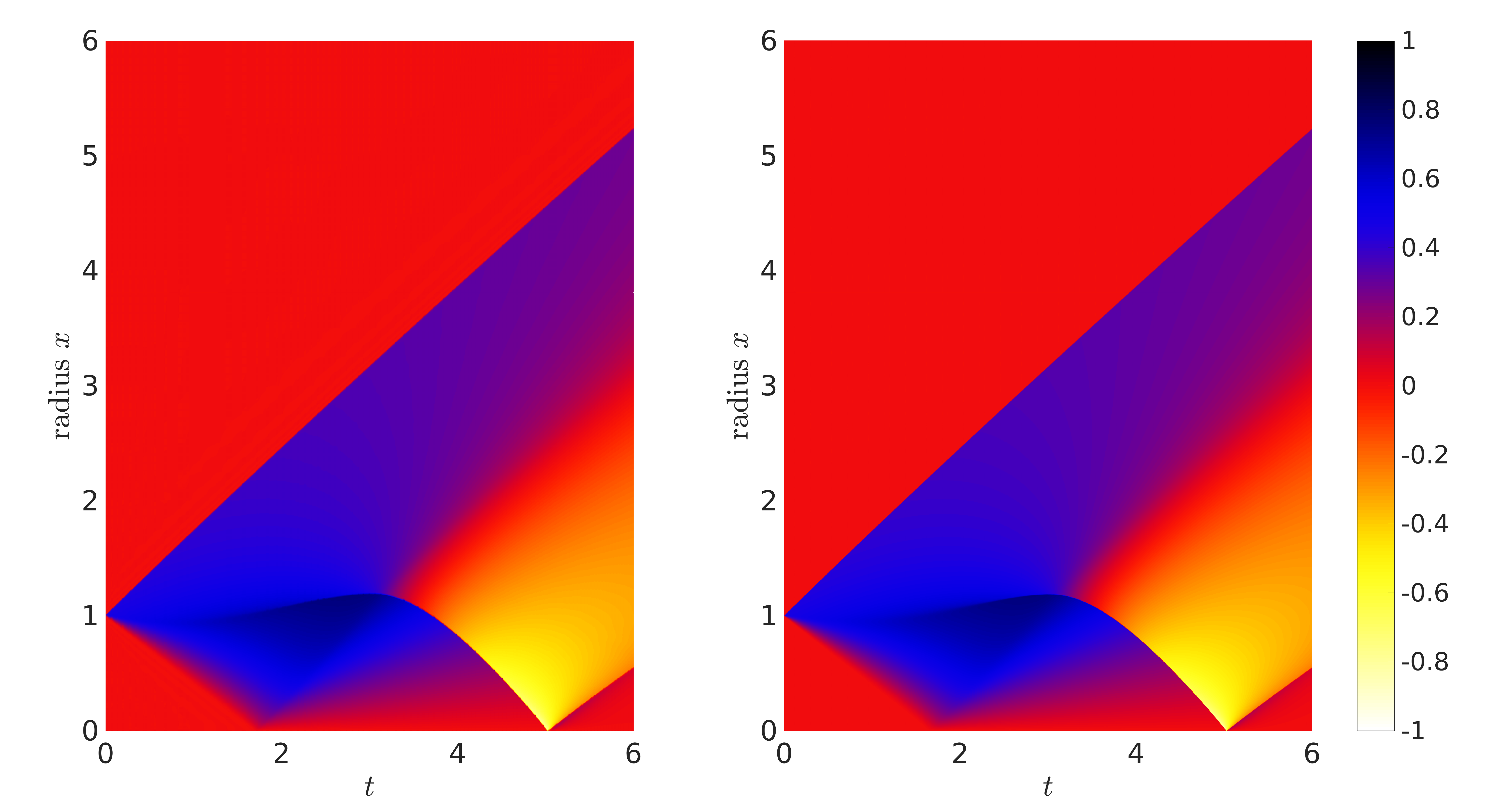}}
    %
    %
%
%
    %
    \subfigure[Results for the pressure $p$]{
    \includegraphics[width=0.55\textwidth]{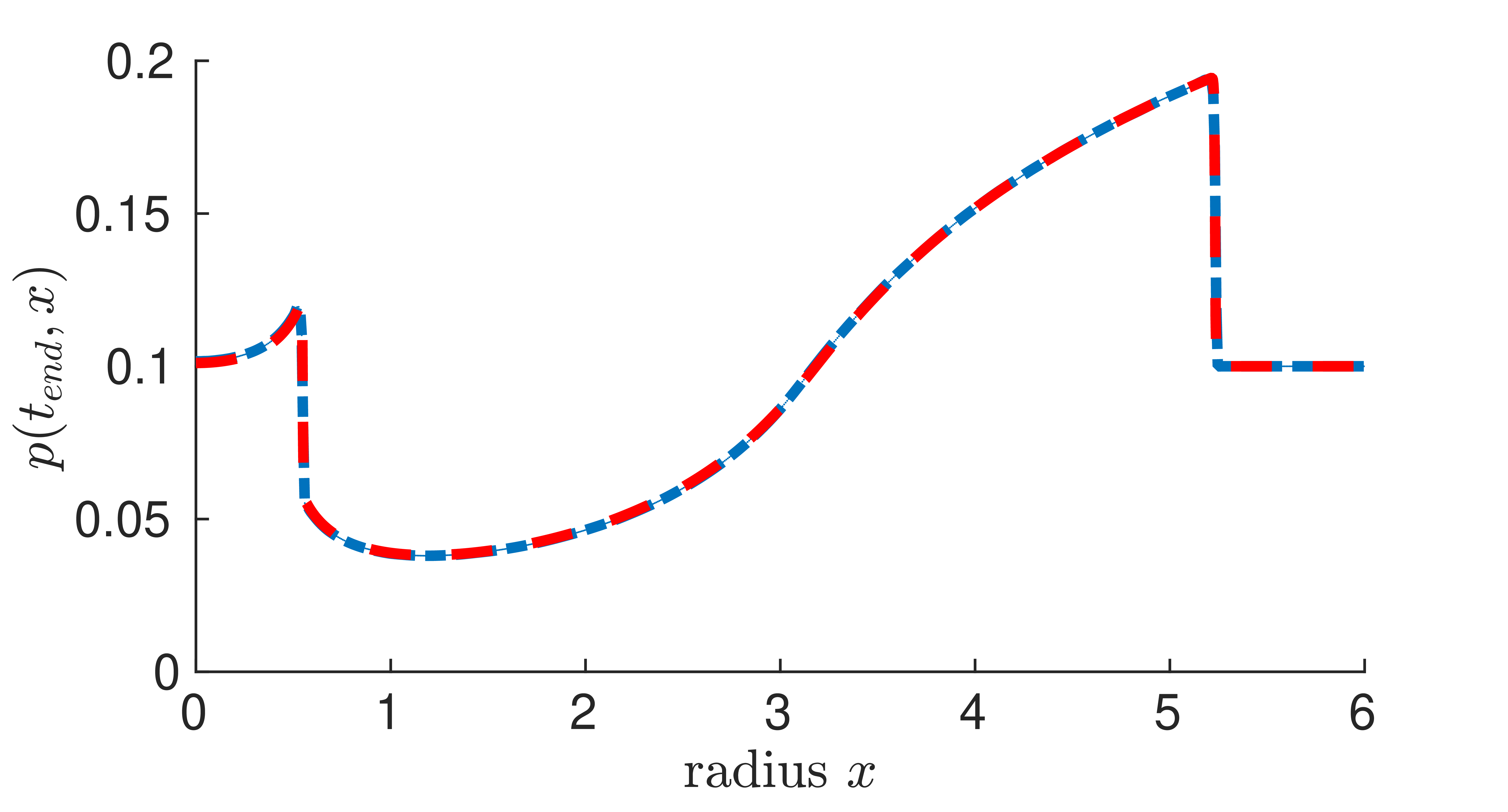}}
    \hspace{-20mm}
    \subfigure[Results for the velocity $v$]{
    \includegraphics[width=0.55\textwidth]{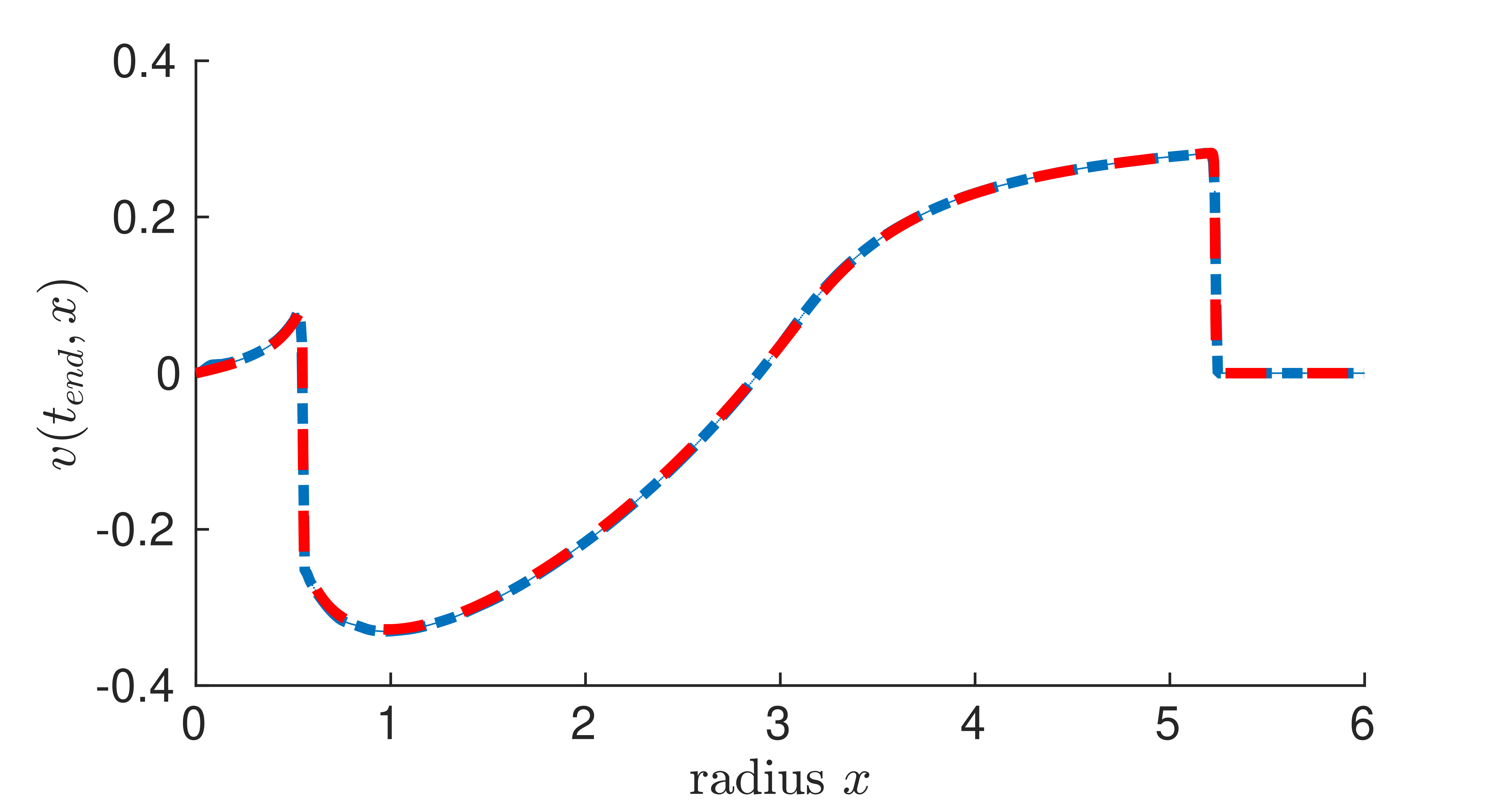}}
    %
    \caption{\textbf{Example 3:} 
    Comparison for MultiWave (left) and RadSymS (right) in the $t$--$x$ plane, see (a), (b).
    Further RadSymS (red) and MultiWave (blue) are compared in radial direction at final time $t_{end} = 6$, see (c), (d), for pressure $p$ and velocity $v$.}
    \label{fig:example-3}
\end{figure}

\begin{figure}[h!]
    \subfigure[Results for the pressure $p$]{
    \includegraphics[width=\textwidth]{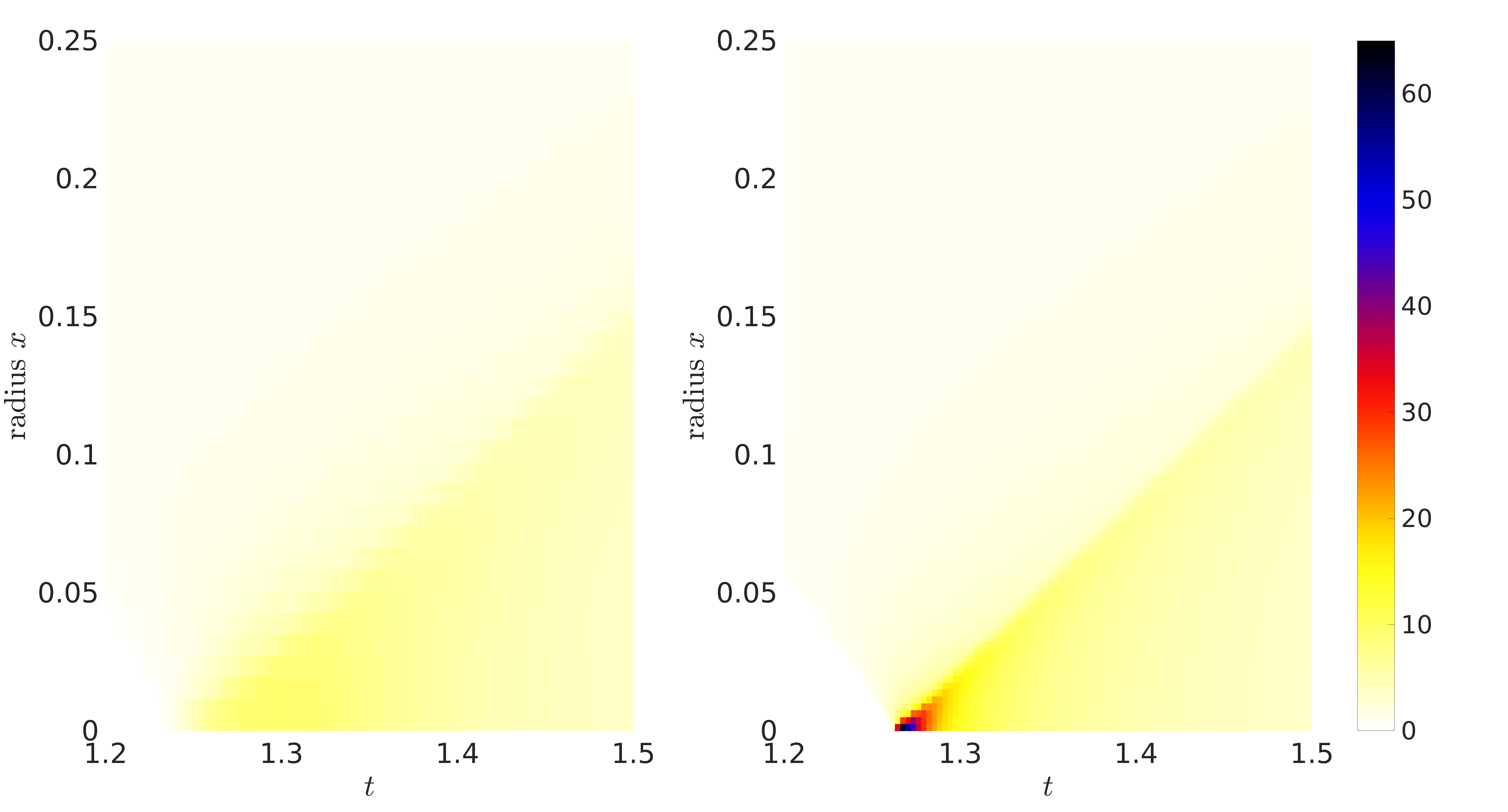}
    }\\
    \subfigure[Results for the velocity $v$]{
    \includegraphics[width=\textwidth]{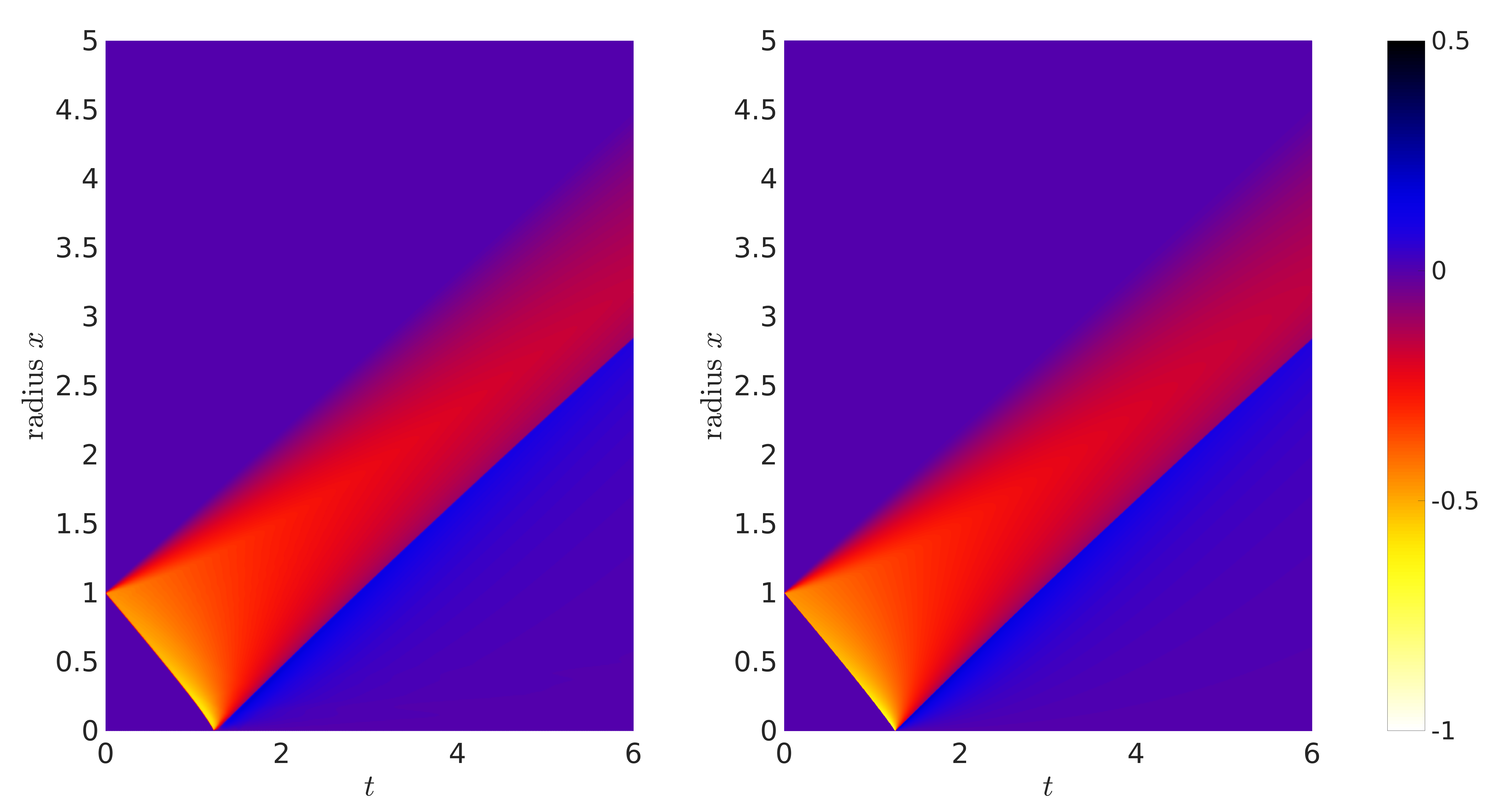}}
    %
    %
%
%
    %
    \subfigure[Results for the pressure $p$]{
    \includegraphics[width=0.45\textwidth]{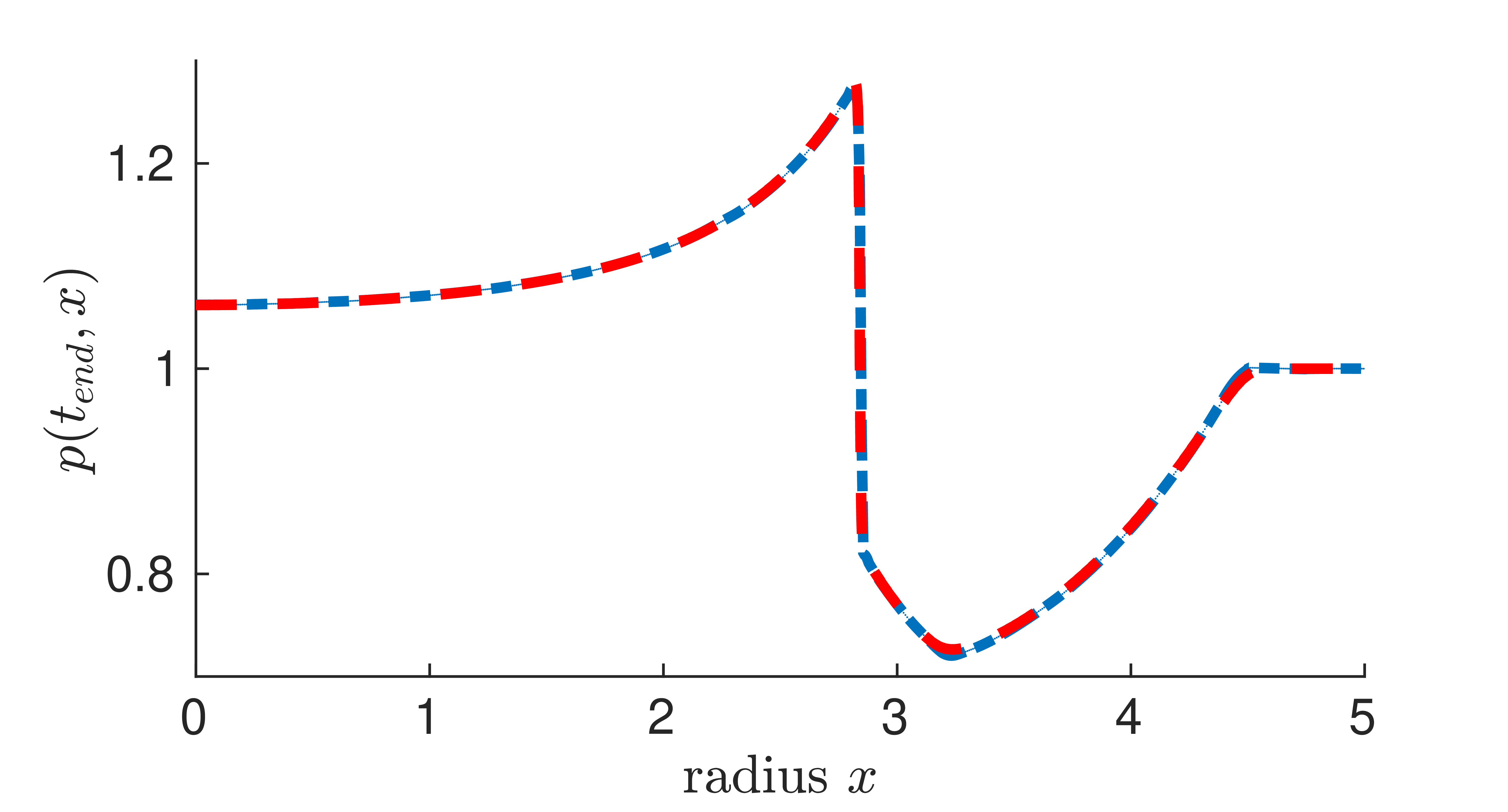}}\quad
    \subfigure[Results for the velocity $v$]{
    \includegraphics[width=0.45\textwidth]{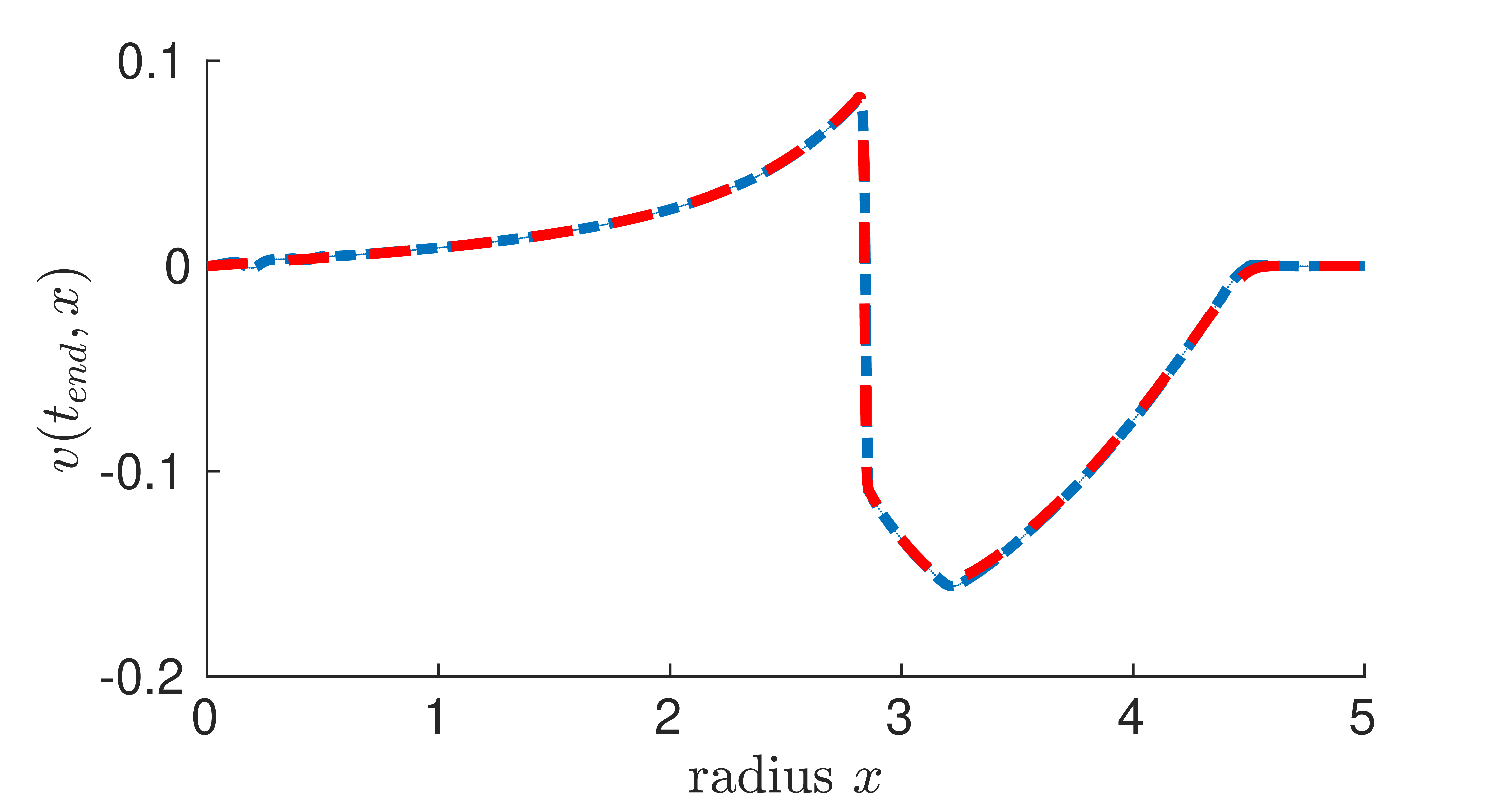}}
    %
    \caption{\textbf{Example 4:} 
     Comparison for MultiWave (left) and RadSymS (right) in the $t$--$x$ plane, see (a), (b).
    Further RadSymS (red) and MultiWave (blue) are compared in radial direction at final time $t_{end} = 6$, see (c), (d), for pressure $p$ and velocity $v$.}
     \label{fig:example-4}
\end{figure}

\begin{figure}[h!]
    \subfigure[Results for the pressure $p$]{
    \includegraphics[width=\textwidth]{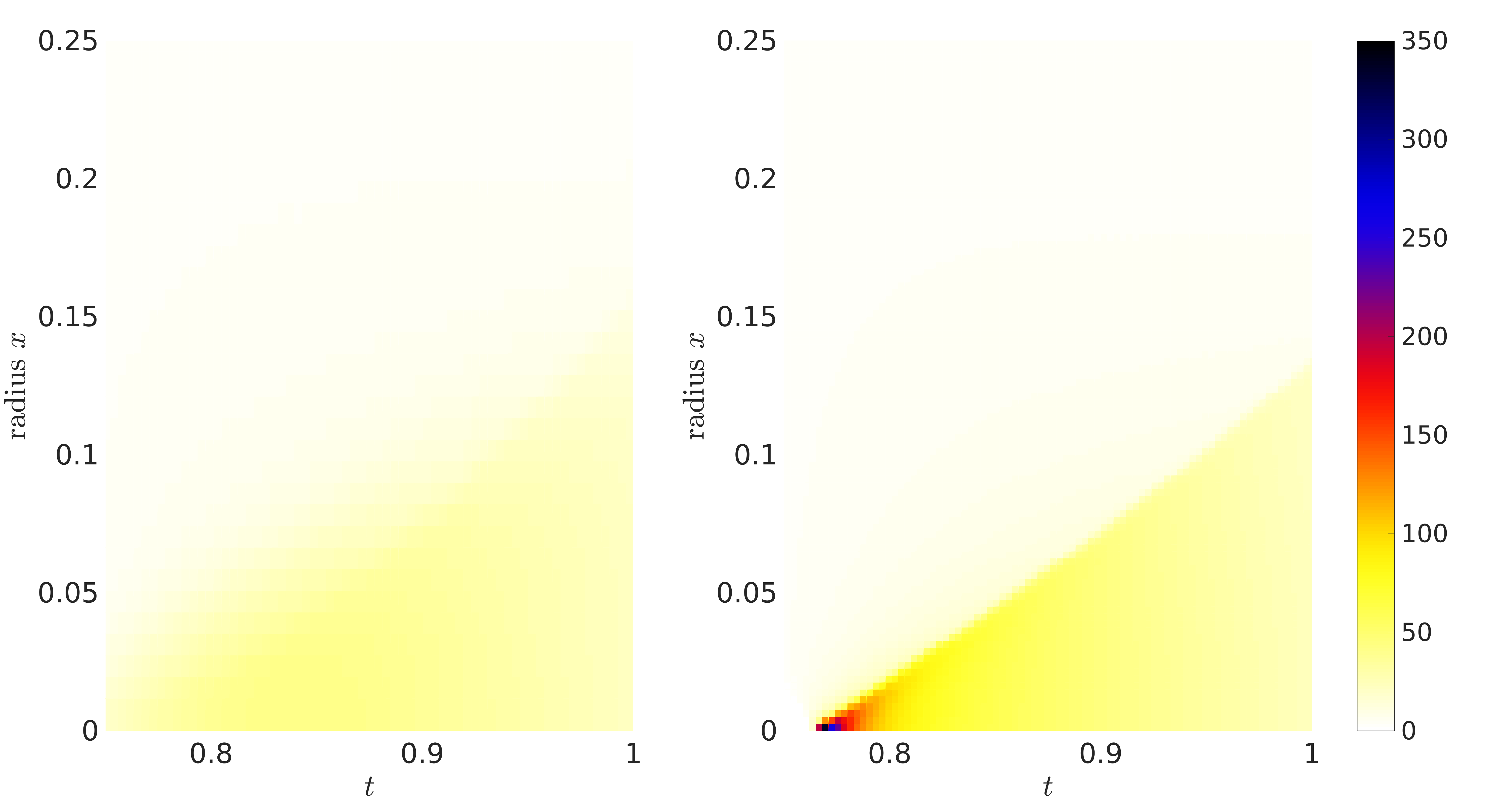}
    }\\
    \subfigure[Results for the velocity $v$]{
    \includegraphics[width=\textwidth]{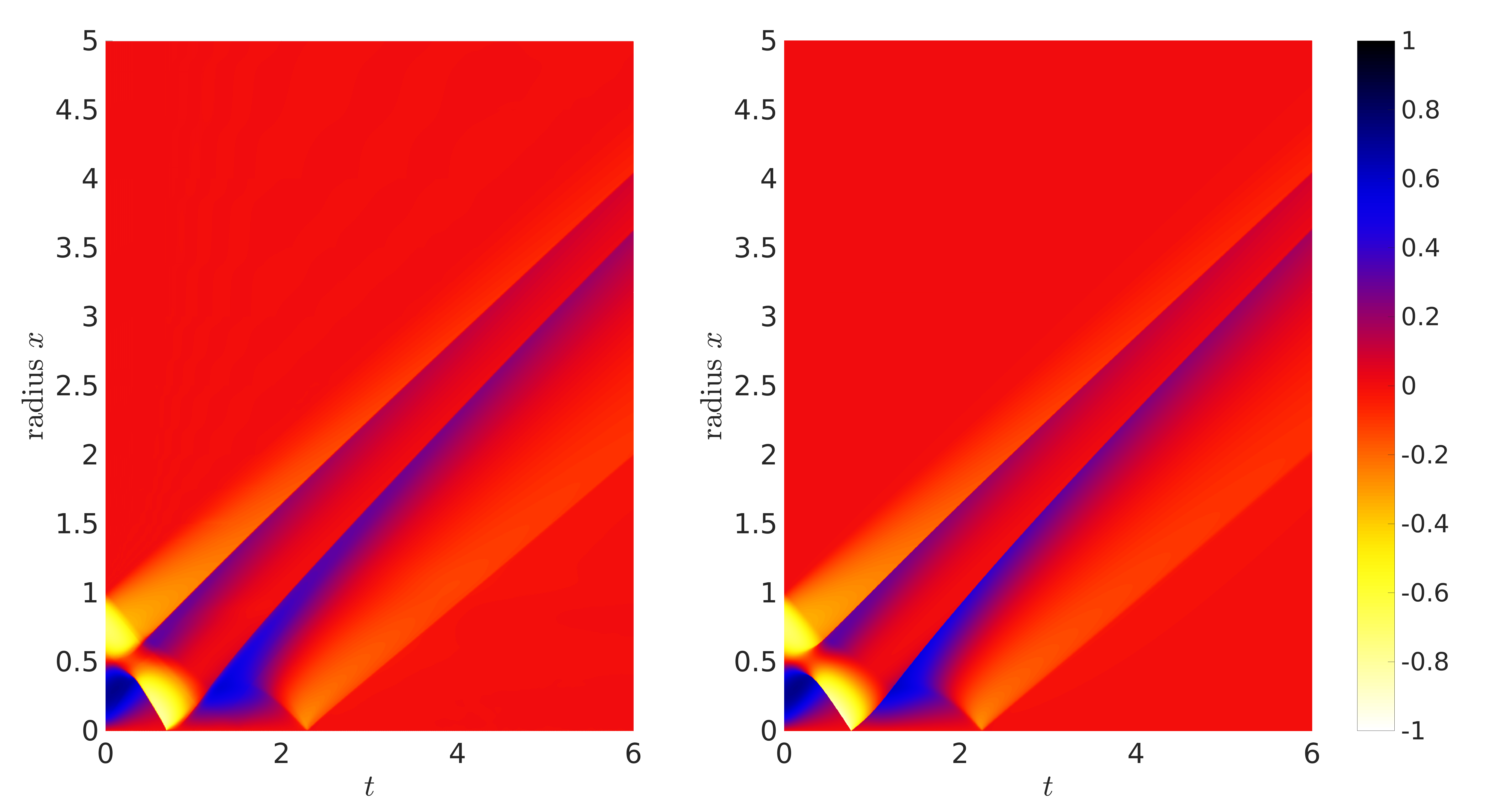}}
    %
    %
%
%
    %
    \subfigure[Results for the pressure $p$]{
    \includegraphics[width=0.475\textwidth]{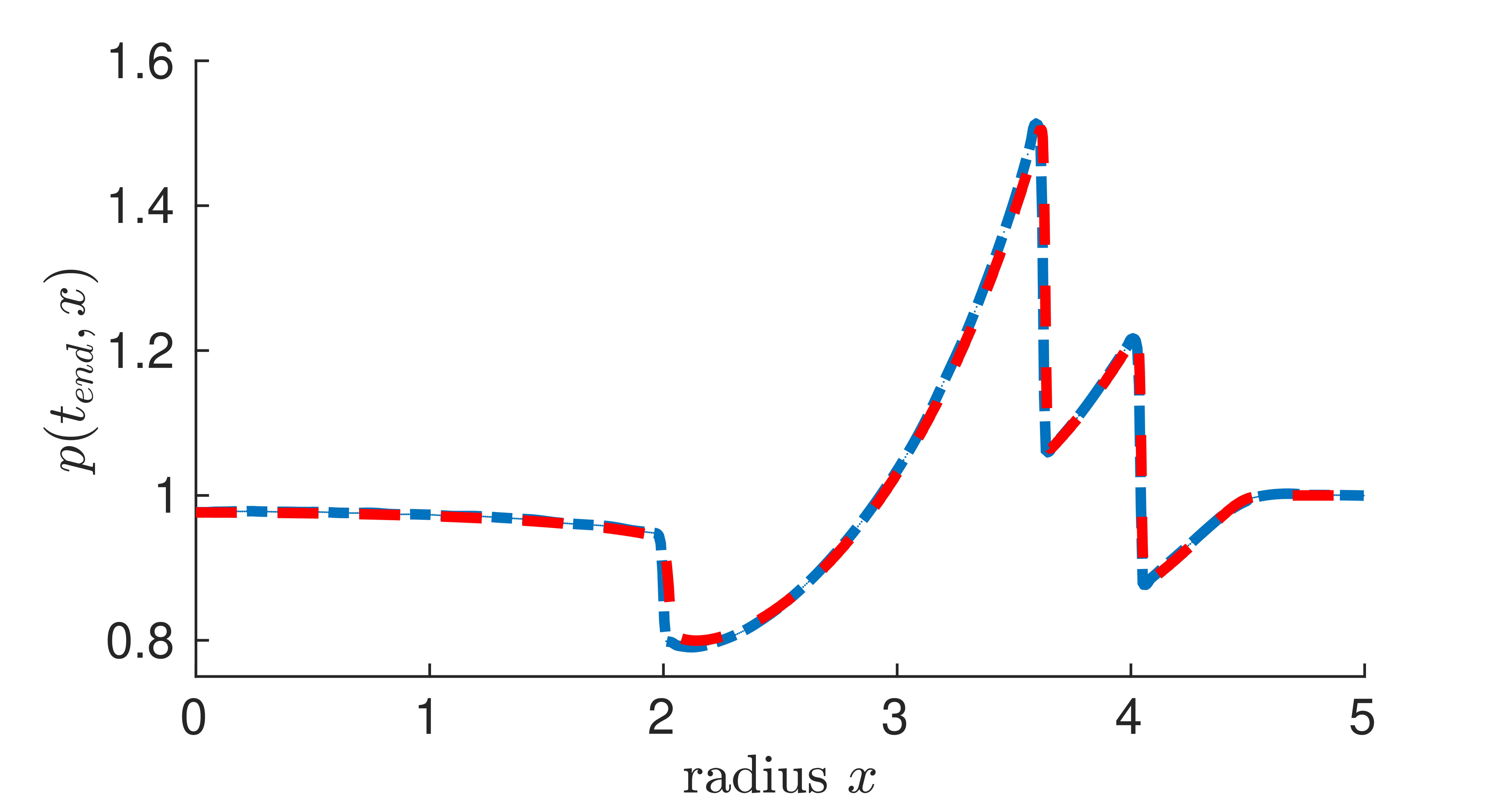}}
    \subfigure[Results for the velocity $v$]{
    \includegraphics[width=0.475\textwidth]{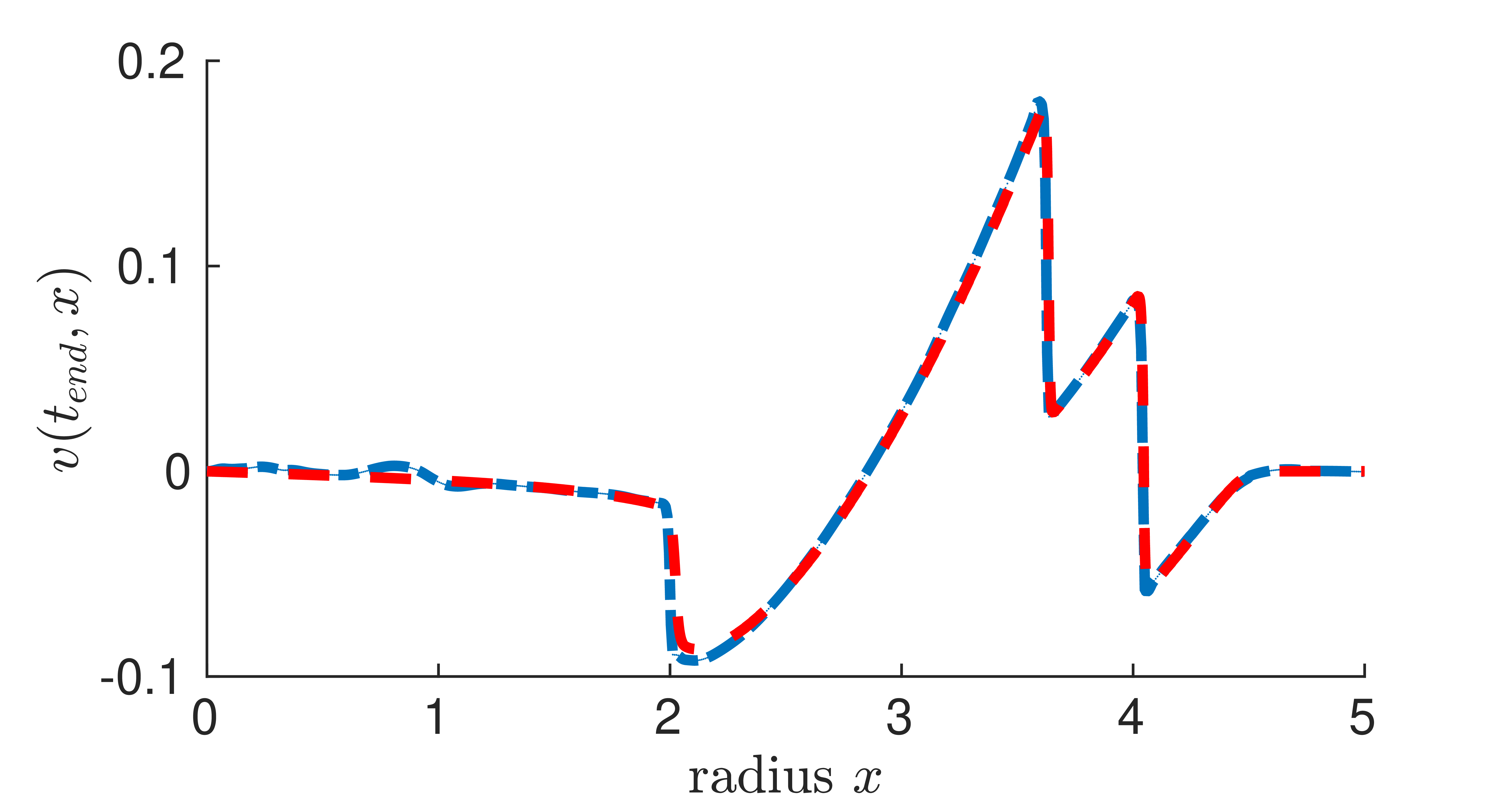}}
    %
    \caption{\textbf{Example 5:} 
     Comparison for MultiWave (left) and RadSymS (right) in the $t$--$x$ plane, see (a), (b).
     Further RadSymS (red) and MultiWave (blue) are compared in radial direction at final time $t_{end} = 6$, see (c), (d), for pressure $p$ and velocity $v$. }
    \label{fig:example-5}
\end{figure}

For all examples we perform computations with the one-dimensional radially symmetric scheme RadSymS and the multi-dimensional DG solver MultiWave for comparisons. Details on the discretization parameters can be found in Sect.~\ref{sec:num_meth}. In Figs.~\ref{fig:example-1} -- \ref{fig:example-5} for Examples 1 -- 5, respectively, 
we present 2D-results in the $t$--$x$ plane and a line plot of the solution in radial direction at the final time $t_{end}$. We compare the results of the two solvers by means of the pressure $p$ and the velocity $v$.
Additionally, for Examples 1 and 2 we compare the solutions also with the reference solution that can be determined for self-similar problems by solving the ODE \eqref{GL_ODE_sys} in the smooth part and using the jump conditions \eqref{shockvalues} to determine the stationary part in Example 1, see Sect.~\ref{configurations}.

\textbf{\emph{Example 1 and 2.}}
The results reported in Figs.~\ref{fig:example-1} and \ref{fig:example-2} clearly show a good agreement of both numerical methods with the reference solution provided by an ODE solver for the final time $t_{end} = 1$ in radial direction. Due to the self-similar structure of the solution we omit the presentation of results in the $t$--$x$ plane.

\textbf{\emph{Example 3.}}
Initially, the pressure inside the bubble is ten times larger than outside, which leads to a fast expansion of the bubble into the outer low pressure area.
This in turn gives rise to the formation of another low pressure area.
We observe the formation of a shock wave, running towards the new low pressure area and reaching the zero point around time $t=5.032$.
The formation of this new shock wave is a peculiar nonlinear phenomenon.
Shortly before the shock reaches the zero point the pressure takes very low values, but its reflection from the zero point depicted in 
Fig.~\ref{fig:example-3} (a) and (b)
indicates a blow up of the pressure in a very small time-space range near the boundary.
This is similar to \cite[Section 5, Example 4]{KLW}, but here the blow up is weaker than in three space dimensions, and its illustration requires a higher numerical resolution.
This may be the reason why this phenomenon was neglected up to day for the explosion test problem with an initial bubble.
We expect a similar behavior for the corresponding solution of the explosion problem with the classical Euler equations.\\
From the depicted solution at time $t=6$ we observe the reflected shock curve at radius around $x=0.55$.
The results in the $t$--$x$ plane given in Fig.~\ref{fig:example-3} (a) and (b) show the expansion of the bubble and the formation of the new shock which focuses at the origin and then is reflected again. At the focus point the pressure rises drastically in a small vicinity of the origin and away from that the pressure values do not vary a lot. Therefore we present a zoom plot of the pressure. We observe an excellent agreement of both numerical methods for the velocity in Fig.~\ref{fig:example-3} (b). For the pressure the results also agree well, except close to the reflection point of the shock at the origin where the DG method cannot resolve this structure due to the resolution of the multi-dimensional method, see Fig.~\ref{fig:example-3} (a). Additionally, we compare both methods 
at final
time $t_{end} = 6$ where the solutions coincide very well, see 
Fig.~\ref{fig:example-3} (c) and (d).

\textbf{\emph{Example 4.}}
The results in the $t$--$x$ plane, see Fig.~\ref{fig:example-4} (a) and (b),
show the collapse of the bubble with a focus point at the origin which then is reflected again. At the focus point the pressure rises drastically in a small vicinity of the origin and away from that the pressure values do not vary a lot. Thus, we again present a zoom plot of the pressure for this example. 
We observe an excellent agreement of both numerical methods for the velocity in Fig.~\ref{fig:example-4} (b). For the pressure the results also agree well except close to the reflection point of the shock at the origin where the DG method cannot resolve this structure due to the resolution of the multi-dimensional method, see Fig.~\ref{fig:example-4} (a).
Additionally, we compare both methods at time $t_{end} = 6$
where the solutions coincide very well, see Fig.~\ref{fig:example-4} (c) and (d).

\textbf{\emph{Example 5.}}
The results in the $t$--$x$ plane, see Fig.~\ref{fig:example-5} (a) and (b),
show a complex wave structure for the velocity and a high pressure focus at around time $t=0.77$. Again, we present a zoom plot of the pressure for this example. We observe an excellent agreement of both numerical methods for the velocity in Fig.~\ref{fig:example-5} (b). For the pressure the results also agree well, except close to the reflection point of the shock at the origin where the DG method cannot resolve this structure due to the resolution of the multi-dimensional method, see Fig.~\ref{fig:example-5} (a). Additionally, we compare both methods at final time $t_{end} = 6$ where the solutions coincide very well, see Fig.~\ref{fig:example-5} (a) and (b).


For all simulations with the multi-dimensional DG solver MultiWave we note some perturbations near the focus point after  wave reflection in this point. These can be observed in the one-dimensional plots for the velocity at the final time $t_{end}$, see in particular
Figs.~\ref{fig:example-1} (b) and \ref{fig:example-5} (d).
Most likely, the perturbations are  caused by a dimensional effect introduced by the Cartesian grids that are not able to preserve and resolve radial symmetry. This dimensional effect can be clearly observed in the adaptive grids not presented here. Under grid refinement the perturbations become smaller and are located more closely to the focus point. Exemplarily, we perform two computations with the DG solver with maximal refinement level $L=8$ and $L=9$ for Example 1. In 
Figs.~\ref{fig:example-1} (a) and (b)
we observe that the perturbation becomes smaller with higher resolution.

\FloatBarrier

\FloatBarrier

\FloatBarrier

\FloatBarrier

\FloatBarrier

\section{Conclusion}\label{sec:conclusion}
In the present work we studied radially symmetric solutions for the ultra-relativistic Euler equations in two and three space dimensions. 
An efficient numerical scheme for the computation of such one-dimensional radially symmetric solutions was developed.\\
Furthermore, we computed solutions for different sets of initial data exhibiting complex wave structures. In the case of self-similar solutions we compared
our method with a reference solution obtained by solving simplified ODEs along a ray $\vartheta = t/x = const$ for fixed time $t_{end}$.
These comparisons show an excellent agreement of the proposed one-dimensional radially symmetric method with the reference solution at some fixed time $t$.\\
By construction one-dimensional solver is capable of calculating radially symmetric solutions very efficiently in comparison to multi-dimensional sol\-vers and, thus, provides reference solutions in the $t$--$x$ plane for comparison with multi-dimensional solutions. Exemplarily, this was demonstrated by performing simulations for the two-dimensional initial value problem applying the DG-solver MultiWave.
These solutions were compared with solutions of the one-dimensional radially symmetric solver. The solutions showed a very good agreement. For this a high resolution was needed for the multi-dimensional DG solver resulting in high computational cost. \\
Finally, we want to emphasize that the examples considered here are very much suited as challenging genuinely multi--dimensional benchmark problems for numerical methods
designed for hyperbolic PDEs. 
The one-dimensio\-nal radially symmetric solver can  provide non-trivial reference solutions at low cost and high accuracy. These reference solutions  may be used for validation of multi-dimensional solvers.

\bigskip

\textit{Acknowledgment.}
The authors thank the Deutsche Forschungsgemeinschaft (DFG, German Research Foundation) for the financial support through  320021702/GRK2326 Energy, Entropy, and Dissipative Dynamics (EDDY), SPP 2410: Complexity, Scales, Randomness (CoScaRa), Projects MU 1422/9-1 (525842915) and TH 2405/2-1 (525939417). 
This project has benefited from funding from the German Federal Ministry of Education and Research (BMBF) through the project grant "Adaptive earth system modeling with significantly reduced computation time for exascale supercomputers (ADAPTEX)" (funding id: 16ME0671).

\newpage
\appendix
\section{Application of Geng Lai's analysis}\label{app:lai}
It is shown by Geng Lai in \cite[Section 2]{GL} that for special solutions depending only on $\vartheta=t/x>0$ the system \eqref{energy_general},
\eqref{momentum_general} can be reduced to the system of ODEs \eqref{GL_ODE_sys}. It is sufficient to discuss
\begin{align}
    \dot{V}(\vartheta) &= (d-1)\frac{V(V-\vartheta)(1-V^2)}{f(\vartheta,V)}\quad\mbox{with} \label{rad_velo_ode}\\
    f(\vartheta,V) &= 3(\vartheta V-1)^2-(V-\vartheta)^2\notag
\end{align}
for $d=2$ and $d=3$ space dimensions, respectively. The differential equation \eqref{rad_velo_ode} has to be supplemented by an initial condition $V(0)=v_0$ with $|v_0|<1$.
For $v_0 \in (0,1)$ we obtain continuous solutions, see \cite[Section 2.2]{GL} for further details. For our discussion we want to focus on the case $-1 < v_0 < 0$ where shock wave solutions occur.
Hence we require an initial condition
\begin{align}
    V(0) = v_0 \in (-1,0)\,. \label{initial}
\end{align}
By $(0,\vartheta_{max})$ with $\vartheta_{max}>0$ we denote the maximal existence interval for the solution $V=V(\vartheta)$ of the initial value problem \eqref{rad_velo_ode}, \eqref{initial}.
It is shown in \cite[Section 2.3]{GL} that $V(\vartheta) < 0$ for $\vartheta \in [0,\sqrt{3}]$ and $v_0 \in (-1,0)$.
Lai obtained that $\sqrt{3} < \vartheta_{max}< \infty$ for the initial condition \eqref{initial}.
Note that we have the normalized speed of light $c = 1$. The values $V=\pm 1$ and $V=0$ are fixed points for the ODE \eqref{rad_velo_ode}.
Hence it is sufficient to study solutions $V=V(\vartheta) \in (-1,0)$ with $\vartheta \in (0,\vartheta_{max})$.
In order to show that \eqref{vvalue} has a unique solution $\tilde{\vartheta} \in (\sqrt{3},\min(3,\vartheta_{max}))$
we first determine the sign of the denominator $f(\vartheta,V)$ on the right-hand side of \eqref{rad_velo_ode}.
Initially we have $f(0,v_0)=3-v_0^2>0$.
As long as $-1<V<0$ we have $V<0$, $V-\vartheta<0$ and $1-V^2 >0$ for the three factors in the nominator of the right-hand side of \eqref{rad_velo_ode}.
We conclude that $f(\vartheta,V)=0$ can only occur at the singular point $\vartheta=\vartheta_{max}$
and obtain by
the intermediate value theorem that $f(\vartheta,V(\vartheta))>0$ and $\dot{V}(\vartheta)>0$ for $0 < \vartheta < \vartheta_{max}$.
Hence $V$ is strictly monotonically increasing in $(0,\vartheta_{max})$. The equation $f(\vartheta,V) = 0$ can be solved explicitly. Reformulating the denominator gives
\begin{align*}
    0 &= 3(\vartheta V-1)^2-(V-\vartheta)^2\\
    &= \left(\sqrt{3}(\vartheta V-1) + (V-\vartheta)\right)\left(\sqrt{3}(\vartheta V-1) - (V-\vartheta)\right)\,.
\end{align*}
We obtain two possible functions where the denominator vanishes, namely $\overline{V}_1, \overline{V}_2 : (\sqrt{3},\infty) \mapsto \R$ with
\begin{align}
    \overline{V}_1(\vartheta) &= \frac{\vartheta + \sqrt{3}}{\sqrt{3}\vartheta + 1} > 0
    \quad\text{and}\quad\overline{V}_2(\vartheta) = -\frac{\vartheta-\sqrt{3}}{\sqrt{3}\vartheta - 1}<0\,.\label{blowup_bound}
\end{align}
Since $V<0$, we can rule out the positive function $\overline{V}_1$.
Due to Lai we obtain from the monotonicity of the solution $V$ that
\begin{equation}\label{monoton}
    \lim_{\vartheta \nearrow \vartheta_{max}}V(\vartheta) = \sup_{0<\vartheta <\vartheta_{max}}V(\vartheta) = \overline{V}_2(\vartheta_{max})\,, \quad
    \lim_{\vartheta \nearrow \vartheta_{max}}\frac{dV}{d\vartheta} = +\infty.
\end{equation}
A straightforward calculation shows for the function $\overline{V}_2$ in \eqref{blowup_bound} that
\begin{equation}\label{ineqv2}
    \overline{V}_2(\vartheta) > \frac{3}{2 \vartheta}-\frac{\vartheta}{2}
\end{equation}
for all $ \vartheta> \sqrt{3}$. The right-hand side of \eqref{ineqv2} vanishes at $\vartheta=\sqrt{3}$ and takes the value $-1$ for $\vartheta=3$.
Using the intermediate value theorem we finally conclude from \eqref{ineqv2} and the first relation in \eqref{monoton} that \eqref{vvalue} has a unique solution
\begin{equation}\label{uniqueshock}
    \tilde{\vartheta} \in (\sqrt{3},\min(3,\vartheta_{max}))\,.
\end{equation}
This guarantees a solution with a shock, see 
Example 1 in Sect.~\ref{configurations}.
According to \eqref{shockvalues} this shock propagates with constant speed $\tilde{s} = 1/\tilde{\vartheta}$.

Estimation \eqref{uniqueshock} contains the unknown quantity $\vartheta_{max}$. Now 
we derive
a better suited explicitly given upper bound for $\tilde{\vartheta}$.
For this purpose we note that the bijective mapping $V_0 : [\sqrt{3},3] \mapsto [-1,0]$ with
\[
  V_0(\vartheta)=\frac{3}{2 \vartheta}-\frac{\vartheta}{2}
\]
has the inverse function $V_0^{-1} : [-1,0] \mapsto [\sqrt{3},3]$ given by
\[
  V_0^{-1}(v)=\sqrt{v^2 + 3}-v\,.
\]
The solution $V$ of the initial value problem is strictly monotonically increasing with $V(0)=v_0$, whereas $V_0$ is strictly monotonically decreasing for $\vartheta > \sqrt{3}$.
Thus, we obtain again from $V_0^{-1}(0) = \sqrt{3}$ and $V_0^{-1}(-1) = 3$ the estimation
\begin{equation}\label{varthetabound}
    \sqrt{3} < \tilde{\vartheta} < \sqrt{v_0^2 + 3}-v_0 < 3\,.
\end{equation}
With \eqref{varthetabound} we have obtained useful bounds for $\tilde{\vartheta}$ and,
hence, for the shock speed $\tilde{s}$ in \eqref{shockvalues} depending only on the initial data but independent of $\vartheta_{max}$.
Moreover, this result holds for $d=2$ as well as for $d=3$ space dimensions.
\newpage
\section{Eigensystem for ultra-relativistic Euler equations}\label{app:eig_sys_ure}
The ultra-relativistic Euler equations can be written in conservative form
\begin{align}
    \label{eq:model}
    \partial_t \bw + \sum_{k=1}^d \partial_{x_k} \bff_k(\bw) = \bzero
\end{align}
for the vector of conserved variables $\bw= (w_1,\ldots,w_d, w_{d+1})=(\obw,w_{d+1})\in\R^{d+1}$ with $d\in\{1,2,3\}$ consisting of  momentum $\obw=(w_1,\ldots,w_d)\in\R^d$ and  energy $w_{d+1}$ and fluxes in the $k$th coordinate direction
\begin{align}
    \label{eq:flux-vec}
    & \bff_k(\bw) = p \be_k + \sum_{i=1}^d \frac{w_i w_k}{w_{d+1}+p} \be_i + w_k \be_{d+1},\
    k=1,\ldots, d,
    %
\end{align}
with $\ \be_i$ the $i$th unit vector in $\R^{d+1}$. This system is closed by an equation of state for the pressure
\begin{align}
    \label{eq;Eos}
    p = p(\bw)
\end{align}
to be specified below.
For its discretization classical finite volume schemes or DG schemes may be applied. For this purpose, numerical fluxes in normal direction $\obn\in\R^d$, $|\obn|=1$,
have to be computed at the interfaces of the elements typically requiring the  eigenvalues and eigenvectors of the corresponding flux Jacobian.
In the following we determine
the eigensystem to the Jacobian of the flux in normal direction
\begin{align}
    \label{eq:flux-normal}
    \bff_\iobn(\bw) := \sum_{k=1}^d \bff_k(\bw) n_k.
    %
\end{align}
The flux Jacobian is determined by
\begin{align}
    \label{eq:flux-normal-jac}
    \bA_\iobn :=\partial_{\ibw} \bff_\iobn(\bw) = 
    \left(
    \begin{matrix}
        \obA   & \oba   \\
        \obn^T & 0
    \end{matrix}
    \right)
\end{align}
with
\begin{align*}
    &  \obA =  \obn \otimes \nabla_\iobw p + \frac{w_n}{w_{d+1}+p} \obI_{d\times d} + \frac{1}{w_{d+1}+p}  \obw \otimes\obn  -\frac{w_n}{(w_{d+1}+p)^2}  \obw \otimes \nabla_\iobw p,\\
    & \oba = \partial_{w_{d+1}} p \, \obn -  \frac{1+\partial_{w_{d+1}}  p }{(w_{d+1}+p)^2}\, w_n \,\obw .
\end{align*}
To compute the eigenvalues we need to determine the characteristic polynomial of the matrix 
\begin{align*}
    %
    \bB_\iobn:= \bA_\iobn - \lambda \bI_{(d+1)\times (d+1)}.
\end{align*}
The derivation is elementary but tedious work performing some algebraic manipulations. It will be helpful to
introduce an orthonormal system $\obt_i\in\R^d$, $i=2,\ldots, d$, to the normal $\obn\equiv \obt_1$ inducing the orthogonal matrix
\begin{align*}
    &\obT_\iobn := (\obt_1=\obn,\obt_2,\ldots,\obt_d),\ \obT_\iobn \obT_\iobn^T = \obI_{d\times d} .
    %
\end{align*}
%
%
Furthermore we introduce the notation
\begin{align*}
    \nabla_\iobn p:= \obT_\iobn^T \nabla_\iobw p,\ \obw_\iobn:= \obT_n^T \obw .
    %
\end{align*}
Then the characteristic polynomial corresponding to the determinant of the matrix $\bB_\ibn$ reads
\begin{align*}
    |\bB_\iobn| &= 
    (-1)^{d+1} \left(\lambda-\frac{w_n}{w_{d+1}+p}\right)^{d-1} p_2(\lambda),
\end{align*}
where the quadratic polynomial $p_2(\lambda)= \lambda^2 + \op \lambda + \oq$ is determined by the coefficients
\begin{align*}
    &\op:=  
    \frac{w_n}{(w_{d+1}+p)^2} \, \sum_{i=1}^{d} \ow_{\iobt_i}  \partial_{\bt_i}  p  -\partial_{\iobt_1}  p    -  \frac{2\,w_n}{w_{d+1}+p} ,\\
    &\oq:= 
    (1 + \partial_{w_{d+1}} p) \frac{w_n^2}{(w_{d+1}+p)^2} -\partial_{w_{d+1}} p  -   \frac{1}{w_{d+1}+p}   \,\sum_{i=2}^{d} \ow_{\iobt_i}   \, \partial_{\bt_i}  p .
\end{align*}
The roots of the characteristic polynomial and, thus, the eigenvalues, are
\begin{align}
    \label{eq:charpol}
    \lambda_0= \frac{w_n}{w_{d+1}+p} \ ((d-1)\text{-multiple}) \quad\text{or}\quad
    \lambda_\pm = 
    -\frac{\op}{2} \pm \sqrt{\left(\frac{\op}{2}\right)^2 - \oq} .
\end{align}
A system of linearly independent right eigenvectors is determined by
\begin{align}
    \label{eq:right-eigenvector-0}
      & \br^i_0 = 
      \bt_i - 
    \alpha_i
    \left( \be_{d+1} +\lambda_0 \bt_1 \right) \in \R^{d+1},\ i=2,\ldots, d, \\
    \label{eq:right-eigenvector-pm}
    &\br_\pm = 
      \lambda_\pm \bt_1 +
    \alpha_\pm
    \sum_{i=2}^d   \,\ow_{\iobt_i} \bt_i  +
                             \be_{d+1} \in \R^{d+1} 
\end{align}
with
\begin{align}
    & \alpha_i := \frac{\partial_{\iobt_i}  p}{\partial_{w_{d+1}} p  +  \lambda_0 \partial_{\iobt_1}  p},\
    & \alpha_\pm :=\frac{w_{d+1}+p}{(w_{d+1}+p)^2 -   w_n^2 }\left(  1   - \frac{w_n}{(w_{d+1}+p)}  \lambda_\pm  \right) .
\end{align}
Here $\bt_i = (\obt_i^T,0)^T$ and  $\bt_1= \bn = (\obn^T,0)^T$.
The system of right eigenvectors is linearly independent, i.e., the matrix
\begin{align*}
    \bR_\iobn = (\br_-,\br_0^1,\ldots,\br_0^{d-1},\br_+)
\end{align*}
is regular. Thus, a system of linearly independent left eigenvectors 
\begin{align*}
    \bL_\iobn=  (\bl_-,\bl_0^1,\ldots,\bl_0^{d-1},\bl_+)^T
\end{align*}
exists that is orthogonal to the right eigenvectors, i.e.,
\begin{align*}
    \bR_\iobn \bL_\iobn = \bI \quad \text{and} \quad  
    \bR_\iobn \bA_\iobn \bL_\iobn = \bLambda_\iobn:=\diag(\lambda_-,\lambda_0 \bone_{d-1}^T,\lambda_+).
\end{align*}
The system of orthogonal left eigenvectors is determined by
\begin{align}
    \label{eq:left-eigenvector-0}
     & \bl_0^i = \frac{1}{D}\left( u_{\iobt_i} (\alpha_+ - \alpha_-) \bt_1 + \sum_{j=2}^d (D \delta_{i,j} -\alpha_j u_{\iobt_i}\beta)\bt_j -  u_{\iobt_i}(\alpha_+ \lambda_- - \alpha_- \lambda_+) \be_{d+1} \right),\  i=2,\ldots,d,\\
    \label{eq:left-eigenvector-pm}
    & \bl_\pm = \mp \frac{1}{D}\left( (1+\alpha_\mp S) \bt_1 + \sum_{i=2}^d \alpha_i(\lambda_0-\lambda_\mp) \bt_i - (\alpha_\mp\lambda_0 S +\lambda_\mp) \be_{d+1} \right)
\end{align}
with
\begin{align}
    & D := \left( \lambda_- + \alpha_- \lambda_0 S \right) \left(1 + \alpha_+ S \right) - \left( \lambda_+ + \alpha_+ \lambda_0 S\right) \left( 1 + \alpha_- S\right),\
    S :=   \sum_{i=2}^d \alpha_i u_{\iobt_i} ,\\
    &  \beta:= \alpha_+ (\lambda_- - \lambda_0) + \alpha_- (\lambda_0 - \lambda_+) .
\end{align}
So far, we have not yet specified an equation of state. For this purpose we rewrite the conserved variables 
in terms of the primitive variables  determined by the velocity $\obu:=(u_1,\ldots,u_d)\in\R^d$ and the pressure $p$, i.e.
\begin{subequations}
    \label{eq:cons-var}
    \begin{align}
        \label{eq:cons-var-a}
        &w_i  := 4 p u_i \sqrt{1 + |\obu|^2},\ i=1,\ldots,d \\
        \label{eq:cons-var-b}
        & w_{d+1}  := p( 3 + 4 |\obu|^2).
    \end{align}
\end{subequations}
Reversely, the primitive variables can be rewritten in conserved variables as
\begin{subequations}
    \label{eq:prim-var}
    \begin{align}
        \label{eq:velo}
        & u_i = \frac{w_i}{\sqrt{4p(w_{d+1}+p)}},\ i=1,\ldots, d,\\
        & p = \frac{1}{3} \left(-w_{d+1} + \sqrt{-3 |\obw|^2 + 4 w_{d+1}^2} \right) \ge 0.
    \end{align}
\end{subequations}
From this we determine the derivatives of the pressure by
\begin{align}
    \label{eq:aux-pres-der-1}
      \partial_{w_j} p =
      \left\{
        \begin{matrix}
       - w_j/(w_{d+1}+ 3p)   &, j=1,\ldots,d \\[2mm]
           (w_{d+1}-p) / (w_{d+1}+ 3p)  &, j=d+1
        \end{matrix}
      \right. .
\end{align}
Introducing normal and tangential momentum $\obw_\iobn= (\ow_{\iobt_1},\ldots, \ow_{\iobt_d})^T$ and velocity $\obu_\iobn= (\ou_{\iobt_1},\ldots, \ou_{\iobt_d})^T$ by means of
\begin{align*}
    &\obw_\iobn:=\obT_n^T \obw = 4 p \obu_\iobn \sqrt{1 + |\obu|^2},\ 
    \ow_{\iobt_i} = \obt_i\cdot \obw,\quad \ow_{\iobt_1} = \bt_1\cdot\obw \equiv \bn\cdot \obw = w_n,\\
    &\obu_\iobn:=\obT_n^T \obu,\ \ou_{\iobt_i} = \obt_i\cdot \obu ,\quad 
    \ou_{\iobt_1} = \bt_1\cdot\obu \equiv \bn\cdot \obu = u_n,
\end{align*}
the pressure gradient and its directional derivatives in normal and tangential direction  are determined by
\begin{align*}
    & \nabla_\iobw p = -\frac{\obw}{u_{d+1}+ 3p} 
    = -2\frac{ \sqrt{1 + |\obu|^2}}{3+2|\obu|^2} \obu ,\
    \partial_{w_{d+1}} p = \frac{w_{d+1} - p}{w_{d+1} + 3 p} = \frac{1 + 2 |\obu|^2}{3 + 2 |\obu|^2},\\
    & \nabla_\iobn p := \obT_n^T \nabla_\iobw p =  
    -\frac{\obw_\iobn}{w_{d+1}+ 3p}  
    = -2\frac{ \sqrt{1 + |\obu|^2}}{3+2|\obu|^2} \obu_\iobn,\\
    &  \partial_{\bt_i}  p = \obt_i \cdot \nabla_{\obw} p =
    - \frac{\ow_{\iobt_i}}{w_{d+1}+ 3p}
    = -2\frac{ \sqrt{1 + |\obu|^2}}{3+2|\obu|^2} \ou_{\iobt_i} .
\end{align*}
Then the eigenvalues in primitive variables read 
\begin{align}
    \label{eq:lambda-spec-pm}
    \lambda_\pm = 
    \frac{2 u_n^2 }{3+2|\obu|^2} \pm  \frac{\sqrt{2(|\obu|^2 - u_n^2)+3}}{3+2|\obu|^2},\
    \lambda_0 = 
    \frac{ u_n }{\sqrt{1+|\obu|^2}} .
\end{align}
The left and right eigenvectors \eqref{eq:left-eigenvector-pm}, \eqref{eq:left-eigenvector-0} and \eqref{eq:right-eigenvector-pm}, \eqref{eq:right-eigenvector-0}, respectively, can be written in primitive variables with
%
%
\begin{align*}
    & w_{\iobt_i} = 4 p \ou_{\iobt_i} \sqrt{1 + |\obu|^2},\
     \alpha_i =
    -\frac{2 \ou_{\iobt_i} \sqrt{1+|\obu|^2}}{1+ |\obu|^2 + \sum_{i=2}^2 (\ou_{\iobt_i} )^2},\ i=2,\ldots,d,\\
    & \alpha_\pm 
      = \frac{1}{4p\left(1+|\obu|^2 -    u_n^2 \right) }\left(  1   - \frac{ u_n }{\sqrt{1+|\obu|^2}}  \lambda_\pm  \right), \
     S =   
        \sum_{i=2}^d \alpha_i 4 p \ou_{\iobt_i} \sqrt{1 + |\obu|^2}.
    %
\end{align*}

\newpage
\bibliographystyle{plain}

\end{document}